\documentclass[twocolumn]{aastex631}

\usepackage{apjfonts}
\usepackage{float}
\usepackage{commath}
\usepackage{amsmath,amssymb}
\usepackage{mathtools}
\usepackage{url}
\usepackage{graphicx}
\usepackage{enumitem}
\usepackage{comment}
\usepackage{multirow} 
\DeclareSymbolFont{matha}{OML}{txmi}{m}{it}
\DeclareMathSymbol{\varv}{\mathord}{matha}{29}

\newcommand\ngc{NGC~7731 }%
\newcommand\msol{$M_{\odot}$}%
\newcommand\rsol{$R_{\odot}$}%
\newcommand\kcsm{$K_\mathrm{CSM}$}%
\newcommand\rcsm{$R_\mathrm{CSM}$}%
\newcommand\Kcsm{K_\mathrm{CSM}}%
\newcommand\Rcsm{R_\mathrm{CSM}}%
\newcommand\Mcsm{M_\mathrm{CSM}}%
\newcommand{\hal}{H$\alpha$}%
\newcommand{\hbe}{H$\beta$}%
\newcommand{\feiio}{\ion{Fe}{2}}%
\newcommand{\oiii}{[\ion{O}{3}]}%
\newcommand{\sii}{[\ion{S}{2}]}%
\newcommand{\nii}{[\ion{N}{2}]}%
\newcommand{\sodiumi}{\ion{Na}{1}}%
\def\simgt{\lower.5ex\hbox{$\; \buildrel > \over \sim \;$}}%
\def\simlt{\lower.5ex\hbox{$\; \buildrel < \over \sim \;$}}%
\def\kspn{KSP-SN-2022c}%
\def\atname{AT~2022ozg}%
\def\ni56{${\rm{^{56}Ni}}$}%
\def\mni56{$M_{\rm{Ni}}$}%
\def\co56{${\rm{^{56}Co}}$}%
\def\dm15{$\Delta$$M_{B,15}$}%
\def\ergs{\rm erg s$^{-1}$}%
\def\t0{$t_{\rm 0}$}%
\def\x1{$x_1$}%
\def\kms{$\rm km\;s^{\rm -1}$}%
\def\vi{\mbox{$V\!-\!i$}}%
\def\zvalue{0.041}%
\def\mhun{mag~(100~days)$^{-1}$}%
\def\mhunn{mag per 100 days}%
\def\mfif{mag~50~days$^{-1}$}%
\def\sfif{$s50_V$}%

\shorttitle{Luminous, Transitional Type II Supernova}
\shortauthors{Jiang et al.}

\begin{document}

\title{Infant Core-collapse Supernovae with Circumstellar Interactions from KMTNet I: 
Luminous Transitional Case of \kspn}
\author{Nan Jiang}
\correspondingauthor{Nan Jiang}
\email{steven.jiang@mail.utoronto.ca}

\author{Dae-Sik Moon}
\affiliation{David A. Dunlap Department of Astronomy and Astrophysics, University of Toronto, 50 St. George Street, Toronto, ON M5S 3H4, Canada}

\author[0000-0003-3656-5268]{Yuan Qi Ni}
\affiliation{Kavli Institute for Theoretical Physics, University of California, Santa Barbara, 552 University Road, Goleta, 93106-4030, CA, USA} 
\affiliation{Las Cumbres Observatory, 6740 Cortona Drive, Suite 102, Goleta, 93117, CA, USA}
\affiliation{David A. Dunlap Department of Astronomy and Astrophysics, University of Toronto, 50 St. George Street, Toronto, ON M5S 3H4, Canada}

\author[0000-0001-7081-0082]{Maria R. Drout}
\affiliation{David A. Dunlap Department of Astronomy and Astrophysics, University of Toronto, 50 St. George Street, Toronto, ON M5S 3H4, Canada}

\author[0000-0002-3505-3036]{Hong Soo Park}
\affiliation{Korea Astronomy and Space Science Institute, 776, Daedeokdae-ro, Yuseong-gu, Daejeon, 34055, Republic of Korea}

\author[0000-0001-9541-0317]{Santiago Gonz\'alez-Gait\'an}
\affiliation{Instituto de Astrof\'isica e Ci\^encias do Espaço, Faculdade de Ci\^encias, Universidade de Lisboa, Ed. C8, Campo Grande, 1749-016 Lisbon, Portugal
}

\author[0000-0001-9670-1546]{Sang Chul Kim}
\affiliation{Korea Astronomy and Space Science Institute, 776, Daedeokdae-ro, Yuseong-gu, Daejeon, 34055, Republic of Korea}
\affiliation{Korea University of Science and Technology (UST), Daejeon 34113, Republic of Korea}

\author[0000-0002-6261-1531]{Youngdae Lee}
\affiliation{Department of Astronomy and Space Science, Chungnam National University, Daejeon 34134, Republic of Korea}

\author[0009-0002-4314-1607]{Ernest Chang}
\affiliation{David A. Dunlap Department of Astronomy and Astrophysics, University of Toronto, 50 St. George Street, Toronto, ON M5S 3H4, Canada}

\begin{abstract}
We present $BVi$ multi-band high-cadence observations of 
a Type II supernova (SN) \kspn\ from a star-forming galaxy at $z$ $\simeq$ 0.041
from its infant to nebular phase. 
Early light curve fitting with a single power-law is consistent with 
the first detection of roughly 15 minutes after shock breakout. 
The SN light curves feature  a rapid rise and decline  across its luminous 
($V$ $\simeq$ --18.41 mag) peak together with a short plateau.
The presence of the short plateau and rapid post-peak decline 
place the SN within a small group of transitional type between Type II-P and II-L subtypes. Its (i) broad and asymmetric H profiles with large emission-to-absorption ratios and (ii) near-peak luminosity in excess of predictions from SN shock cooling models both point to circumstellar interactions in this SN. Early colour evolution exhibits a short-lived blueward motion in \bv\ within 
the first few days and continuous reddening in \vi, inconsistent with simple blackbody heating. 
Our simulations of SN light curves estimate 13 \msol\ and 680 \rsol\ for the mass and radius of the progenitor, respectively, together with CSM of 0.73 \msol\ to account 
for the excess luminosity and rapid post-peak declines. 
We discuss the origin of its short plateau and early colour evolution in the context of partial envelope stripping of the progenitor star and a delayed SN shock breakout near the edge of the CSM, respectively, 
as indicated by our simulations. 
We establish a correlation between post-peak decline rates and CSM mass in Type II SNe, 
highlighting that CSM interactions play a major role in shaping the post-peak
evolution of transitional types.
\end{abstract}


\section{Introduction}

Hydrogen-rich or Type II supernovae (SNe) are known to result from the
core collapse of massive ( $\simgt$ 8\msol) stars \citep{Colgate_66_progenitior, Smart_09_progenitior, smarrt_15_image}. Their 
first observable electromagnetic signature is the shock breakout (SBO) when light escapes from the ejecta \citep{Ohyama_63_SBO, SNEC_16_early}. 
Early light curves of H-rich SNe are dominated by 
shock cooling emission (SCE), where the shock-heated material dissipates energy through radiation. 
After the light curves reach the peak, 
those with a large H envelope enter a plateau phase maintained by H recombination in their post-peak decline, 
known as Type II-P SNe \citep{Popov_93_Henv}. 
The evolution during the plateau phase depends on various properties such as the mass and radius
of the progenitor as well as the explosion energy \citep{Popov_93_Henv, Kasen_09_Henv, Kepler_16, Fang_24_Henv}. 
Some H-rich Core Collapse Supernovae (CCSNe) exhibit a linear post-peak decline without the apparent presence
of a plateau, forming a group of Type II-L SNe \citep{Filippenko_97_SNclass}.

The light curve differences between Type II-P and II-L events 
are often attributed to differences in 
H envelope mass of progenitors 
\citep[e.g.,][]{Heger_03_Henv, Eldridge_18_binary}, 
although pre-explosion images of CCSN progenitors 
have identified that both types originate from cold supergiants \citep[e.g.,][]{smarrt_15_image, VanDyk2017_progen, VanDyk_24_image, Xiang_24_progenitior}.
A growing number of studies 
has suggested that the two subtypes are the extreme ends of a single 
continuous distribution rather than being intrinsically separate classes 
\citep[e.g.,][]{Anderson_14_sample,Sanders_15_population,Dessart_24_binarycsm}.

A continuous distribution for Type II-P and II-L is 
further supported by the identifications of a few transitional 
cases---such as SNe 2006Y, 2006ai, 2013by, and 2016ezg \citep{Valenti_16_class, Hiramatsu_21_physical}---that
straddle the subtypes by exhibiting a plateau followed 
by a rapid post-peak decline.
Most of the transitional cases show signatures of circumstellar material (CSM) 
interactions 
revealed by narrow emission lines at early phases ($\simlt$ 3 days) from shock ionization or 
excess emission \citep[e.g.,][]{Valenti_16_class, Hiramatsu_21_physical}.
Both transition cases and II-L SNe are typically more 
luminous than Type II-P at peak due to
contributions from CSM interactions 
\citep[e.g.,][]{Gall_15_risetime, Valenti_16_class, SNEC_17_CSM,SNEC_18_CSM},
supporting that CSM interactions play an important role 
in the transition between the subtypes. 
Transitional events have also been observed to have relatively 
shorter plateaus compared to typical Type IIP SN \citep{Hiramatsu_21_physical}. 
This is compatible with them having smaller H envelopes at the time of explosions, which has been interpreted to be the result of envelope stripping in some cases \citep{Hiramatsu_21_physical}.

The configuration of CSM, its interactions with the ejecta, 
as well as mass loss and potential envelope stripping 
in the transitional events 
have been poorly understood mainly because of the lack 
of observed events---only a handful number of transitional 
events have been identified to date 
as far as we are aware \citep[e.g.,][]{Valenti_16_class, Hiramatsu_21_physical}. 
Early observations before the peak are critical to 
identify and investigate transitional SNe because 
the CSM interactions
take place during early phases of SN explosions 
\citep{Ginzburg_14_CSM,Piro21,Kozyreva_22_massloss}.

In this study, we present the discovery and analyses of an infant 
Type II SN (\kspn),
exhibiting transitional behaviour between Type II-P and II-L subtypes
featured with a large peak luminosity, rapid evolution, 
and evidence for strong CSM interactions.
In \S\ref{sec:method}, we present our photometric and spectroscopic observations of \kspn\ 
as well as our analysis of the host galaxy.
We conduct light curve analyses in \S\ref{sec:analysis}, 
including classification, colour evolution, 
and \ni56\ mass estimation.
Our spectrum of \kspn, which is presented in \S\ref{sec:spec},
confirms its Type II nature alongside several signatures originating from CSM interactions.
We present light curve modelling in \S\ref{sec:modeling} 
before we provide a discussion on 
the transitional nature and progenitor system of \kspn\ in \S\ref{sec:discuss}
and summary and conclusion in \S\ref{sec:conclude}.

\section{Observations and Data Analysis}\label{sec:method}

\subsection{Photometry and Light Curves}\label{sec:pho}

We conducted high-cadence, multi-band observations on a $2\degr \times 2\degr$ field 
containing the nearby galaxy \ngc\ in $BVI$ bands as part of the Korea Microlensing Telescope Network (KMTNet) Supernova program \citep{Moon_16_ksp,2021Moon}. 
The KMTNet consists of three identical 
1.6 m telescopes located in Australia, South Africa and Chile, providing 24-hour
continuous monitoring of the sky with a 0\farcs4 per pixel sampling \citep{Kim_16_ksp}.
We obtained about 1500 60-s images
reaching a 3$\sigma$-detection limit in the range of 21--22 mag 
in each of $BVI$ bands between 2020  and July 2022 December
when the field was observable. 
The typical cadence between observations in the same filter is $\sim6$ hours.

\kspn\ was first detected with a S/N ratio of $\sim 4$ at 2 hours and 33 minutes on 2022 July 13 (UTC, or MJD = 59773.102) as a new point source  at the location of (RA, decl) = (${\rm 23^h43^m44.47^s,  -38\degr30\arcmin55\farcs44}$) (J2000)
with an apparent magnitude of 20.26 $\pm$ 0.37 mag in the $B$ band.\footnote{\kspn\ was also reported as \atname\ by the Asteroid Terrestrial-impact Last Alert System (ATLAS; \citealt{2022ozg})
at the location (RA, decl) = ($\rm 23^h43^m44.47^s, -38\degr 30\arcmin 55\farcs36$) with an apparent magnitude of 18.25 mag in the orange band
at 2022, July 18 (UT).} 
It was subsequently detected in two 60-s $V$-band images 
obtained between 2 and 3 hours later than the first $B$-band detection. 
We bin these two adjacent $V$-band images and adopt
their mean epoch as the epoch of the first $V$-band detection,
and apply the same procedures to the two first I-band detections. 
The first $V$- and $i$-band magnitudes are 
20.25 $\pm$ 0.15 and 20.79 $\pm$ 0.15 mag, respectively. 
(Note that although the KMTNet $I$-band filter is similar to the standard
Johnson $I$-band filter \citealt{Kim_16_ksp}, 
our photometric calibration is made against $i$-band magnitudes
of standard stars in the American Association of Variable Star Observers 
[AAVSO] catalogue.
As a result, the magnitudes of \kspn\ are given in $i$-band magnitudes.)

\begin{figure*}
\centering\includegraphics[width=\linewidth]{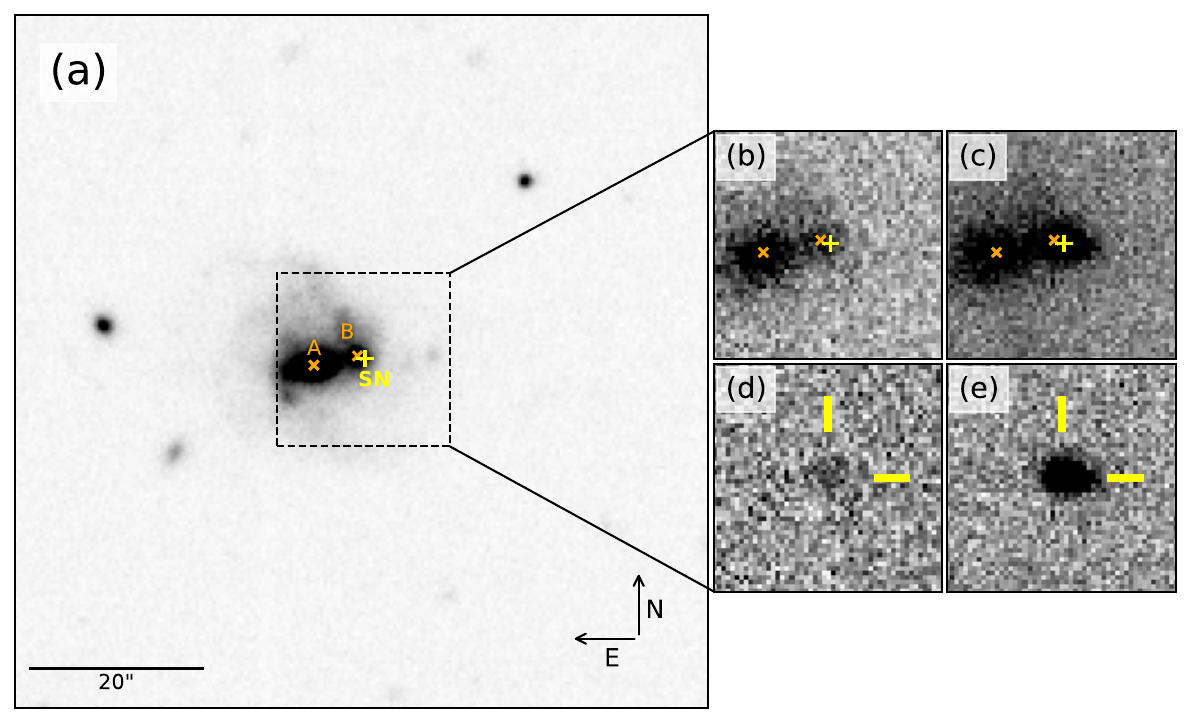}
    \caption{KSP images centred at the location of \kspn. 
    (a) $B$-band pre-SN explosion image created by stacking 340 60-s exposures 
    obtained before July 2020 showing the extended host galaxy around the center.  
    The yellow cross marks the location of \kspn\ at the western end of the host galaxy
    that appears to be mainly composed of two components A and B  (two yellow symbols of 'x'). 
    Their coordinates are  (RA, decl) = (${\rm 23^h43^m44.98^s, -38\degr30\arcmin56\farcs13}$) 
and (${\rm 23^h 43^m 44.55^s, -38\degr 30\arcmin 55\farcs10}$) (J2000) for A and B, respectively.
    North is up and east is to the left.
    (b) The 20\arcsec $\times$ 20\arcsec  $I$-band discovery image of 
    \kspn\ at 03:24:27 on 2022 July 13 (UTC).
    (c) Same as (b) but for the image with the peak $I$-band brightness
    at 01:59:25 on 2022 July 23.
    (d) Same as (b) but after subtraction of a pre-SN template image, 
    showing the first appearance of \kspn.
    (e) Same as (c) but after subtraction of a pre-SN template image.}
    \label{fig:big}
\end{figure*}

\autoref{fig:big}(a) is a pre-SN $B$-band image centred on the location of \kspn\ 
created by stacking 340 60-s exposures obtained before 2022
to reveal its surroundings and the host galaxy. 
The SN location (yellow cross) is near 
the western end of an extended source which we identify to be 
the host galaxy of \kspn\ (see \S\ref{sec:spec_reduction}).
We apply two photometric methods in our estimation of SN magnitudes:
image subtraction and multi-object PSF (MOP) fitting method, 
both implemented in the custom Python-based software 
SuperNova Analysis Package (SNAP)\footnote{\href{https://github.com/niyuanqi/SNAP}{https://github.com/niyuanqi/SNAP}} 
that we have developed for the photometry 
of the KMTNet data \citep[][]{SNAP, Ni_22_method,Ni_23_method,Ni_MOP}.
In our photometry based on image subtraction method,
we first create template images for each band  by stacking several individual exposures
obtained before the first detection of the source,
and subtract the template images from the target exposures 
using the High Order Transform of PSF ANd Template Subtraction package \citep[HOTPANTS;][]{HOTPANTS}.
We then conduct aperture photometry on subtracted images with a 
Kron radius \citep{kron} obtained from fitting a Moffat function \citep{Moffat}
to about 10 nearby isolated AAVSO standard stars\footnote{ The AAVSO Photometric All-Sky Survey: Data Release 9, https://www.aavso.org/apass}.
The image subtraction is very sensitive to the seeing 
and the presence of CCD artefacts in a given image. As a result,
we are only able to obtain acceptable subtraction results for $\sim$ 60\% 
of the images.
\autoref{fig:big} (d) and (e) show the subtracted images
of \kspn\ for its first detection and peak brightness, respectively,
in the $I$ band.

\autoref{fig:big} shows that the galaxy located to the east of the SN 
is extended and several nearby clumps can be identified,  
particularly those labelled as $A$ and $B$.
In our photometry based on the MOP fitting technique \citep[see Appendix A of ][for the details of the MOP fitting]{Ni_MOP},
we first choose to model $A$ and $B$ as two independent 
PSF-convolved S{\'e}rsic profiles in the absence of a better model to 
describe the light profile for the extended/clumpy galaxy. 

We estimate S{\'e}rsic profile parameters for $A$ and $B$ using images created by stacking several pre-SN exposures with a reduced $\chi^2$ value $\simeq$ 1.
The final step of our MOP-based photometry for \kspn\ is simultaneously 
fitting the source together with $A$ and $B$ in our science images.
For this, we use a Moffat function to model \kspn\ 
using the fixed PSF parameters obtained from fitting the same function 
to about 10 nearby AAVSO standard stars
and the fitted S{\'e}rsic parameters for $A$ and $B$.
The free-fitting parameters in this process are 
three scaling parameters for the SN, $A$, and $B$
as well as two additional parameters for modelling the background. 

\begin{figure}
    \centering
    \includegraphics[width=\linewidth]{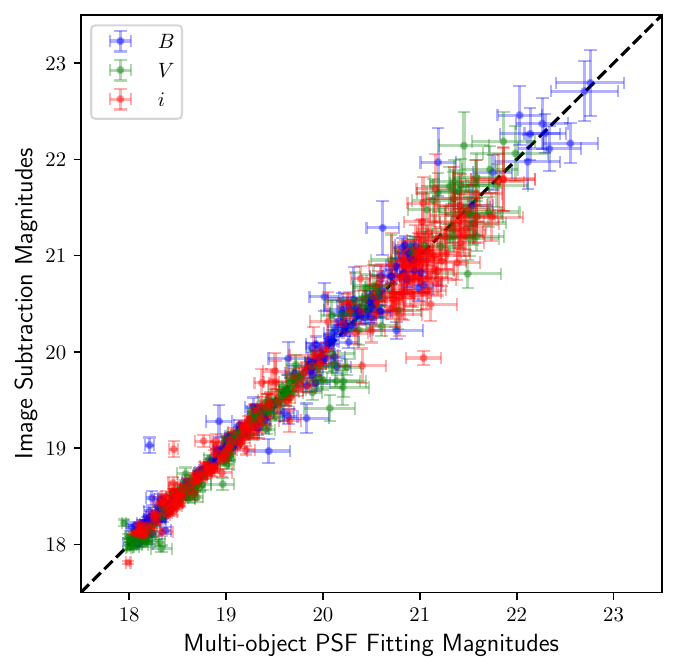}
    \caption{Comparison between magnitudes of \kspn\ obtained from the MOP fitting method (abscissa) and image subtraction method (ordinate) for the $B$ (blue), $V$ (green) and $i$ (red) bands. 
    The black dashed line represents the fitted linear relation between them
    with a slope of $\sim 1$}
    \label{fig:comp}
\end{figure}

\autoref{fig:comp} provides comparisons between the magnitudes of \kspn\ 
obtained by the image subtraction and MOP fitting methods,
showing a good agreement between them with the fitted slope (dashed line) of $\sim 1$.
Our final magnitudes of \kspn\ consist of those from the image subtraction method,
which is about 60\%, with the remaining part from the MOP fitting method.

\begin{figure*}
    \centering
    \includegraphics[width=\linewidth]{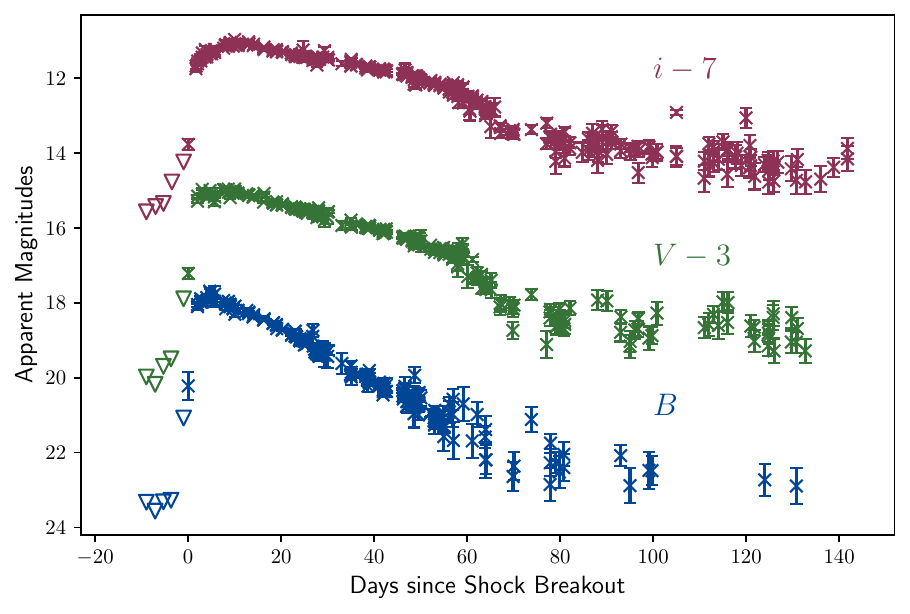}
    \caption{The observed apparent magnitudes (crosses with error bars)
    of \kspn\ in $B$ (blue), $V$ (green), and $i$ (red) bands.
    The open triangles represent the upper limits for detection at S/N = 3.
    The $V$ and $i$ band magnitudes are offset by 3 and 7 mags, respectively.}
    \label{fig:raw}
\end{figure*}

We convert the KMTNet $BVI$ instrumental magnitudes 
to the standard Johnson $BV$- and Sloan $i$-band magnitudes 
using 10 AAVSO reference stars near \kspn. 
Because of the transmission difference  between the KMTNet and Johnson 
$B$-band filters, 
the calibration of the KMTNet $B$-band data against the Johnson $B$-band
magnitudes of the AAVSO references require a colour correction of
$\Delta B$ $\simeq$ 0.27 (\bv) + offset \citep[see][for details]{Park_17_ksp}, 
where $\Delta B$ is the $B$-band magnitude differences between
the KMTNet magnitudes before colour correction and the standard 
magnitudes of the AAVSO references from the database. 
We implement this by correcting the $B$-band magnitudes 
of the AAVSO references used in our photometric calibration 
using their known \bv\ colours. 
No such colour dependence has been identified in the KMTNet $V$- and $i$-band
magnitudes \citep{Park_17_ksp, Park_19_ksp}. 
Table~\ref{tab:photometry} and \autoref{fig:raw} present our photometric measurements for \kspn\ .

\begin{deluxetable}{ccc}
\tabletypesize{\footnotesize}
\tablecolumns{3} 
 \tablecaption{\kspn\ Apparent Magnitudes}\label{tab:photometry}
 \tablehead{
 \colhead{MJD (days)} & \colhead{Magnitude $\pm$ Uncertainty} & 
 \colhead{$$B$$and}
 } 

\startdata 
59772.104 & $<$ 21.077 & $B$ \\
59772.105 & $<$ 20.894 & $V$ \\
59772.107 & $<$ 21.241 & $i$ \\
59773.102 & 20.258 $\pm$ 0.370 & $B$ \\
59773.127 & 20.246 $\pm$ 0.149 & $V$ \\
59773.128 & 20.787 $\pm$ 0.149 & $i$ \\
59774.749 & 18.768 $\pm$ 0.091 & $i$ \\
59774.793 & 18.695 $\pm$ 0.095 & $i$ \\
59775.089 & 18.142 $\pm$ 0.081 & $B$ \\
59775.090 & 18.317 $\pm$ 0.052 & $V$ \\
59775.092 & 18.637 $\pm$ 0.037 & $i$ \\
59775.134 & 18.084 $\pm$ 0.062 & $B$ \\
59775.135 & 18.182 $\pm$ 0.038 & $V$ \\
59775.137 & 18.566 $\pm$ 0.034 & $i$ \\
59775.741 & 18.525 $\pm$ 0.065 & $i$ \\
59775.782 & 17.932 $\pm$ 0.078 & $B$ \\
59775.783 & 18.020 $\pm$ 0.084 & $B$ \\
59775.788 & 18.558 $\pm$ 0.047 & $i$ \\
... & ... & ...
\enddata
 \tablecomments{Sample $BVi$ magnitudes of \kspn\ without extinction nor $K$-correction.
 This table is published in its entirety in the electronic edition. A
portion is shown here for guidance regarding its formatting.}
\end{deluxetable}

According to the Milky Way extinction map from \citet{SF11}, 
the $E$(\bv) value in the direction of \kspn\ is 0.0107 mag.
This requires reddening correction of 0.045, 0.033, and 0.021 mag
for the $BVi$ magnitudes at the location of the SN, respectively, 
based on the reddening law from \citet{Fitzpatrick_extinction} with $A_V/E(B-V)=3.1$. 
In the absence of \sodiumi\ D $\lambda$5890, 5896 feature in 
our spectrum of the host of \kspn\ (see \autoref{fig:spec_all}),
no correction for the host extinction is applied. 
We perform $K$-correction for the SN magnitudes 
using the redshift $z$ $\simeq$ \zvalue (see \S\ref{sec:spec_reduction} for the
redshift determination).
Given the low redshift of the source, we only take the $1+z$ term 
into account \citep{Kcorr} as opposed to utilizing information on 
the spectral shape of the source.
\autoref{fig:raw} shows the light curves of \kspn\ after
the extinction and $K$-correction (see \S\ref{sec:lc} for the analysis
of the light curves).

\subsection{Spectroscopy and Host}\label{sec:spec_reduction}

We obtained spectra of \kspn\ using the GMOS spectrograph \citep{Hook_2004}
on the Gemini South 8~m telescope on 
2022 September 2 and 3 when the SN was at 55--56 days 
since the $B$-band peak brightness.
We first obtained three red spectra of 2 $\times$ 300s exposure
using the R400 grating for the three central wavelengths of 7370, 7600, and 7820 \AA\
to combine them to create the final red spectrum.
We applied the same set-up of three 2 $\times$ 300s exposures 
for our blue spectrum  but with the B600 grating and 
the central wavelengths of 4900, 5200, and 5500 \AA. 
The 1\arcsec\ slit was placed at a position angle of 96\degr to obtain spectra of both the SN and host simultaneously.
We adopted the \texttt{gmos} suite of \texttt{pyraf} tasks \citep{1993ASPC.52.173T} 
for our spectroscopic data reduction,
including bias subtraction, flat fielding, 
wavelength calibration using CuAr lamps,
skyline subtraction, 
flux calibrations with the spectrophotometric standards CD-32 9927 (R400) and LTT 7379 (B600),
and spectrum extraction.

\begin{figure*}
    \centering    
    \includegraphics[width=\linewidth]{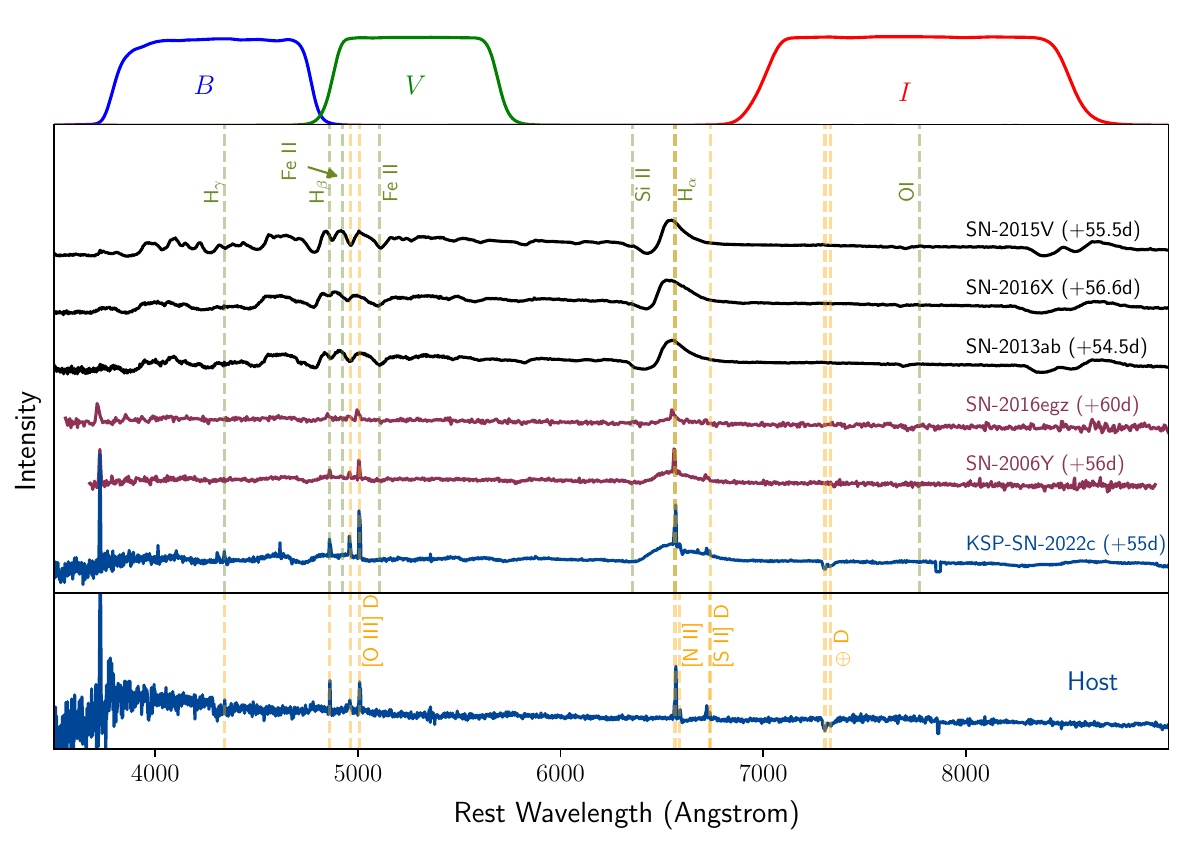}
    \caption{Comparisons between the spectra of \kspn\ and its host with
    those of five Type II SNe obtained at similar epochs: three normal Type II-P events
    SN~2013ab, SN~2015V, and SN~2016X \citep{Jaeger_19_berkerly}
    and two short-plateau, transitional events SN~2006Y and SN~2016egz \citep{Hiramatsu_21_Halpha}. 
    The KMTNet filter transmission curves are shown at the top of the plot.}
    \label{fig:spec_all}
\end{figure*}

\autoref{fig:spec_all} compares our Gemini spectra of \kspn\ and its host 
with those of five Type II SNe  observed at similar epochs \citep{Hiramatsu_21_physical, Jaeger_19_berkerly},
identifying \hal\ and \hbe\ from \kspn\ and its host alongside some other spectral features.  
The \hal\ emission of \kspn\ is notably broad with a narrow peak
which is also present in the host spectrum 
extracted from the sources A and B (\autoref{fig:big}) to the east of \kspn\ 
along the slit positioned with a PA of 96\degr.
The central wavelengths of the narrow peak of \hal\ and also \hbe\ emission
in the host spectrum obtained with Gaussian fitting 
are 6832.4 and 5058.9 \AA, respectively, corresponding to a redshift $z$ $\simeq$ \zvalue.
The sources A and B together is known as LCRS B234107.3--384735 \citep{Shectman_96_galaxy},
and we identify it, which is a barred spiral galaxy, 
to be the host galaxy of \kspn\ based on their positional overlap, 
identical redshifts, and star-forming activities of the galaxy. 
The host spectrum in \autoref{fig:spec_all} shows the presences of \oiii\ ($\lambda\lambda$ 4963 5006), 
\nii\ ($\lambda$ 6585), and \sii\ ($\lambda\lambda$ 6736 6738).
Our penalized PiXel-Fitting \citep[pPXF,][]{Cappellari_17_galaxy} analysis using 
E-MILES stellar population model \citep{Vazdekis_16_galaxy} of the host spectrum 
finds the host to be young with a luminosity weighted age of 0.49--0.89 Gyr 
and a metallicity not significantly different from the solar value
at a redshift of $\simeq$ 0.041. 
Its near ultra-violet magnitude $NUV_0$ $\simeq$ 18.47 from 
GALEX observations \citep{Martin_05_galaxy} corresponds to a star formation 
rate of 0.2 \msol\ yr$^{-1}$ \citep{Kennicutt_98_starformation}.
We estimate the stellar mass of the host galaxy to be  9.1 $\times$ 10$^8$ \msol\
based on its brightness of 16.97, 17.01, and 16.58 in the
$B$, $V$, and $i$ bands, respectively \citep[see Table 7 in ][]{Bell_03_galaxy}.
The presence of strong spectroscopic features of H and the star-forming
host galaxy confirm that \kspn\ is a Type II SN.
The details our spectroscopic analyses, comparisons, and
interpretations are presented in \S\ref{sec:spec}.

In the frame of the standard $\Lambda$CDM cosmology with a Hubble constant of 
67.4 \kms\  Mpc$^{-1}$, matter density parameter 0.315, 
and vacuum density parameter 0.685 from the Planck measurements \citep{Plank20},
we obtain 188 Mpc and 36.4 mag as the luminosity distance
and distance modulus for \kspn.

\section{Light Curve Analysis}\label{sec:analysis}

\subsection{Light Curve Evolution and Classification}\label{sec:lc}

\autoref{fig:raw} shows the $BVi$ light curves of \kspn, showing our first 
detection is a few days before the peak brightness. 
The light curves are featured with a rapid rise to the peak and 
a post-peak decline during $\sim$ 60--100 days which is followed by  
a tail phase exhibiting a gradual decline.

\begin{figure}
    \centering\includegraphics[width=\linewidth]{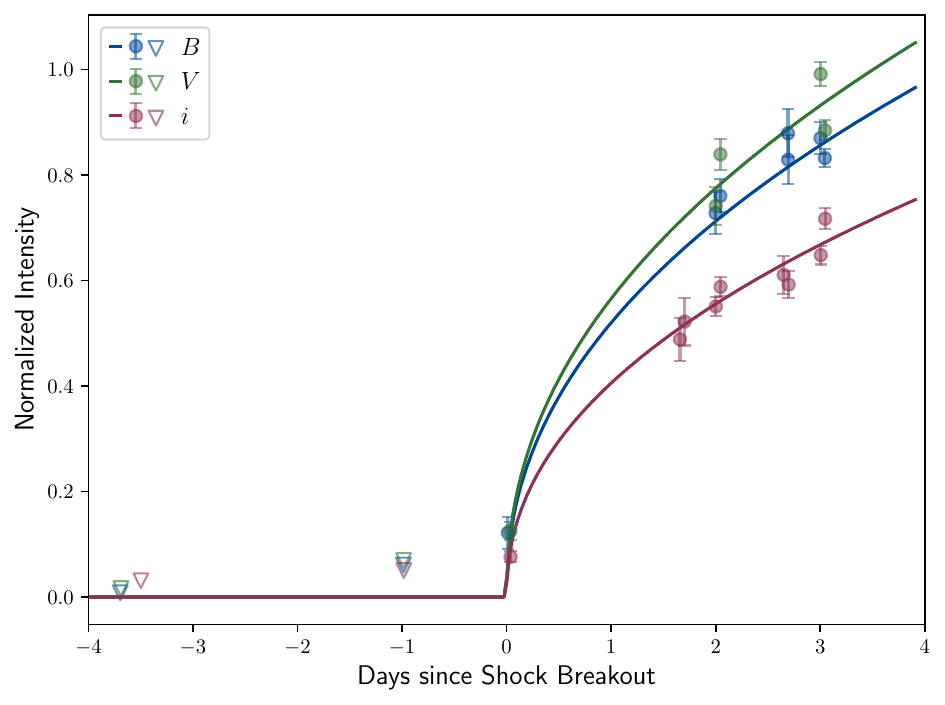}
    \caption{Single power-law fitting (solid curves) of early light curves (filled circles) 
    of \kspn\ normalized by the peak intensities of each band.
    The open triangles represent the upper limits for detection at S/N = 3.
    The best-fit results are $t_0$ = 59773.09 $\pm$ 0.01 days (MJD) for the epoch 
    of SBO and $n$ =  0.45 $\pm$ 0.06 for the power-law index.
    }
    \label{fig:early}
\end{figure}

We estimate the epoch of shock breakout ($t_0$) for \kspn\ 
by fitting the following power-law equation to the early 
($\simlt$ 3.5 days from the first detection) rising part of the light curves:
$f_\lambda = C_\lambda(t-t_0)^n$,
where $f_\lambda$, $C_\lambda$, $t$, and $n$ denote the brightness, scale parameter,
time, and power-law index, respectively \citep[e.g.,][]{Afsariardchi_19_TypeIIobs}.  
\autoref{fig:early} shows the results of our power-law fitting
with the best-fitted parameters 
$t_0$ = 59773.09 $\pm$ 0.01 days 
(or 0.2 $\pm$ 0.1 hours before the first detection at MJD = 59773.102) and
$n$ = 0.45 $\pm$ 0.06
obtained from simultaneous fitting of the $BVi$ light curves.

The fitted power-law index of \kspn\ is smaller than the prediction 
by a homologous expansion in which the luminosity is $\propto t^2$,  
while it is comparable to 0.32 of the infant Type IIP SN of KSP-SN-2016kf \citep{Afsariardchi_19_TypeIIobs} and  
consistent with $\simlt$ 1.5 obtained from a sample of Type II SNe 
powered by SCE \citep[e.g.,][]{Piro_nakar_13,gaitan_15_risetime}.
The fitted SBO epoch, on the other hand, 
which is only 0.2 $\pm$ 0.1 hours before our first detection,
makes \kspn\ a Type II SN with one of the earliest detections from SBO in the optical bands.
The previous earliest detections have been in the range of 1--3 hours, including 
$\sim$ 2 hours for KSN2011a and KSN2011d \citep{Kepler_CCSN}, 
$\sim$ 3 hours for SN~2013fs \citep{Yaron_17_13fs}, 
$\sim$ 2 hours for KSP-SN-2016kf \citep{Afsariardchi_19_TypeIIobs}, and
$\sim$ 1 hour for SN~2023ixf \citep{Wang_ifx}. 
Recent observations of nearby SN~2023ixf and SN~2024ggi with an extremely high cadence 
showed that the rising part of their light curves are better fitted with two power-law
components rather than a single component \citep{Wang_ifx, Shrestha_24_ggi}. 
Given the cadence of our observations of \kspn, it is impossible to conduct 
two power-law component fitting to its early light curves, and we adopt
$t_0$ = 59773.102 (MJD) as the epoch of SBO for the SN from the single component fitting.

We estimate the epochs of the peak brightness of \kspn\ to be 
MJD = 59777.7 $\pm$ 1.1, 59777.9 $\pm$ 1.3, and 59779.8 $\pm$ 1.9 days
(or 4.5, 4.8, and 6.8 days from  $t_0$) in $B$, $V$, and $i$ bands, respectively,
by fitting the general empirical form for SN light curves
from \citet[][see Eqn. 1 therein]{Bazin_09_peak} 
to its light curves (\autoref{fig:raw}) up to 60 days from the first detection.
The rise times \kspn\ between the epoch of SBO and those of fitted peak brightness are 
4.6, 4.8, and 6.7 days in $BVi$, respectively,
considerably shorter than $6.1_{-2.3}^{+2.5}$ ($B$),
$7.9_{-2.9}^{+3.6}$ ($V$), and $10.1_{-2.3}^{+3.6}$ days ($i$) 
obtained from a sample of Type II SNe  \citep{gaitan_15_risetime} 
at the rest frame wavelengths of 4600, 5600, and 8200 \AA, respectively.
These wavelengths roughly correspond to the rest-frame isophotal wavelengths 
of the KMTNet $BVi$ bands for \kspn\  for its redshift (\S~\ref{sec:spec}).
\kspn, therefore, appears to be one of the fast-rising Type II SNe. 
The observed peak magnitudes of \kspn\ are 17.70 ($B$), 17.96 ($V$) and 18.11 ($i$)
which correspond to absolute  magnitudes of --18.52 ($B$), 
--18.41 ($V$), and --18.26 ($i$) mag after reddening correction, respectively.
While it is more luminous than most of Type SNe 
by almost two magnitudes \citep[e.g.,][]{Anderson_14_sample, Jaeger_19_berkerly},
its peak luminosity is  comparable to that of CCSNe with CSM interactions
\citep[][i.e. $\sim -18$ mag]{Jacobson_24_CSM},

The post-peak light curve decline patterns of \kspn\ (\autoref{fig:raw}) 
are different between $B$ band and $V,i$ bands.
The $B$-band light curve shows a simple linear post-peak decline until about 70 days 
after which it becomes flat; in contrast,
the $V,i$-band light curves show three phases in their post-peak decline:
(1) plateau-like declines until about 60 days;
(2) change in the decline patterns during $\sim$ 60--80 days; and
(3) slowly declining ``nebular phase'' at $\sim$ 80 days onward.
The short-duration phase (2) between the plateau and nebula phases
is called the ``drop'' phase serving as evidence for the presence 
of a plateau phase in Type II-P SNe
\citep{Olivares_10_drop,Valenti_16_class, Anderson_14_sample, Valenti_16_class, Zhang_21_Halpha, SN1980K, SN2018gj}.
We estimate the decline rates ($\dot{m}$) for the three phases 
of \kspn\ light curves using the following relations:

\begin{equation}
    m_\lambda = 
    \begin{cases}
    A_1 + \dot{m}_1(t-t_p) & t\leq t_E\\
    A_2 + \dot{m}_2(t-t_E) & t_E<t\leq t_N\\
    A_3 + \dot{m}_3(t-t_N) & t> t_N
    \end{cases}
    \label{eq:threephase}
\end{equation}
where $A$ is the beginning magnitude of each phase and $t$ is time.
The parameters $t_p$, $t_E$, and $t_N$ are the beginning epochs
of the three phases, respectively, 
where $t_p$ is identical to the peak epoch and
$t_N$ is the beginning epoch of the nebular phase.
$t_E$ is exclusively for the $V, i$ bands for their drop phase.
The continuity of the light curves require 
$A_2=A_1+\dot{m}_1(t_E-t_p)$ and $A_3=A_2+\dot{m}_2(t_N-t_E)$. 
\autoref{tab:fit_parm} lists the best-fit values of 
the parameters obtained by fitting \autoref{eq:threephase} to the light curves
of \kspn\  with a reduced $\chi^2$ value $\sim$2. 
In the fitting $t_p$ is fixed to be those values 
obtained for the epochs of peak brightness above.

\begin{deluxetable*}{ccccc}
\tabletypesize{\footnotesize}
\tablecolumns{5} 
\tablewidth{0.99\textwidth}
\tablecaption{Best-fit post-peak decline epochs ($t$) and rates ($\dot{m}$) of \kspn\ 
obtained by fitting \autoref{eq:threephase} to  the $BVi$ (\autoref{fig:raw}) 
and bolometric (\autoref{fig:bol_comp}) light curves (\S\ref{sec:Lum}).
All the epochs are measured relative to SBO.} 
\tablehead{
\colhead{Parameters} & \colhead{$B$} & \colhead{$V$} & \colhead{$i$} & \colhead{Bolometric} 
} 
\startdata
$t_t$ (days) & -- & -- & -- & 29.34 $\pm$ 1.08 \\
$t_E$ (days) & -- & 58.99 $\pm$ 0.80 & 57.20 $\pm$ 0.76 & 50.00 $\pm$  0.82 \\
$t_N$ (days) & 66.21 $\pm$ 3.98 & 68.15 $\pm$ 0.89 & 70.21 $\pm$ 1.03 & 75.14 $\pm$ 2.73 \\
$\dot{m}_t$ (\mhunn) & -- & -- & -- & 1.96 $\pm$ 0.03 \\
$\dot{m}_1$ (\mhunn) & 6.58 $\pm$ 0.06 & 3.39 $\pm$ 0.03 & 2.34 $\pm$ 0.03 & 0.95 $\pm$  0.07 \\
$\dot{m}_2$ (\mhunn) & -- & 16.3 $\pm$ 2.30 & 9.90 $\pm$ 0.90 & 3.92 $\pm$ 0.04 \\
$\dot{m}_3$ (\mhunn) & 1.35 $\pm$ 0.93 & 0.98 $\pm$ 0.21 & 1.16 $\pm$ 0.21 & 0.70 $\pm$ 0.26\\
\enddata
\end{deluxetable*}
\label{tab:fit_parm}

\begin{figure*}
    \centering
    \includegraphics[width=\linewidth]{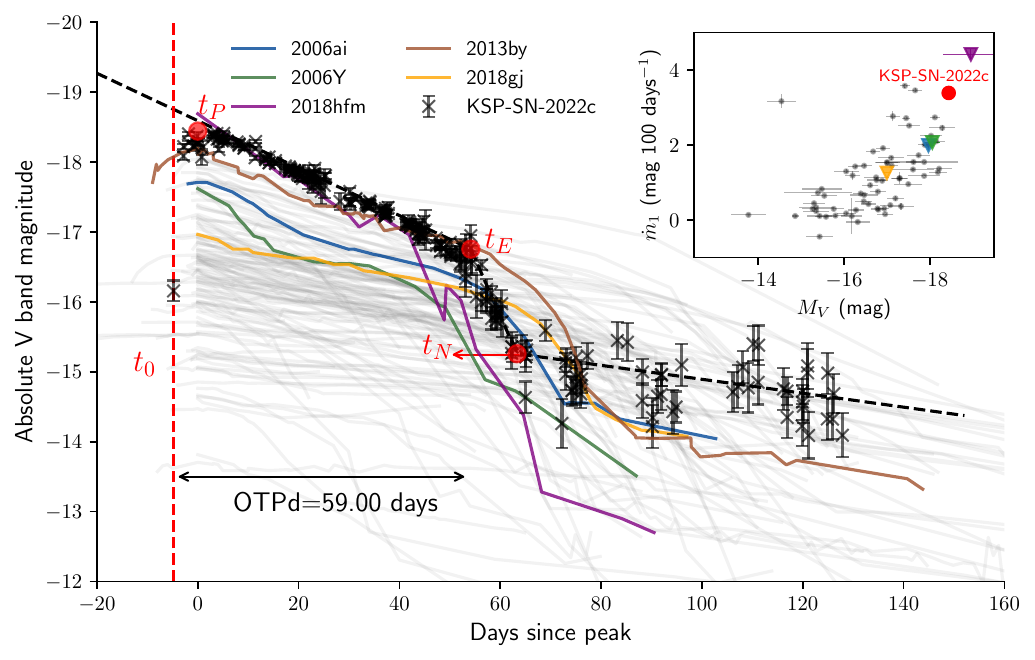}
    \caption{The absolute $V$-band light curve of \kspn\ (black crosses with error bars) compared
    with those of 116 samples of Type II SNe \citep[grey curves;][]{Anderson_14_sample}, 
    SN~2006ai and SN~2006Y \citep[blue and green curves, respectively;][]{Hiramatsu_21_Halpha}, 
    SN~2018gj \citep[orange curve;][]{SN2018gj}, 
    SN~2018hfm \citep[purple curve;][]{Zhang_21_Halpha}, 
    and SN~2013by \citep[sienna curve;][]{Valenti_16_class}. 
    The black dashed lines represent the fitted light curves using
    \autoref{eq:threephase}, while the red vertical dashed line
    marks the epoch of SBO (= $t_0$).
    The red-filled circles mark the three epochs of \autoref{eq:threephase} for \kspn.
    The inset compares the peak $V$-band magnitudes (filled grey circles) 
    and $\dot{m}_1$, which is the decline rate during the plateau phase,
    of the sample Type II SNe from \citet{Anderson_14_sample}.
    The four filled triangles show the locations of SN~2018gj (yellow),
    SN~2006ai (blue), SN~2006Y (green), and SN~2018hfm (purple). 
    The blue and green triangles are partly overlapping. 
    The filled red circle is for \kspn.}
    \label{fig:V_LC}
\end{figure*}

The fitted decline rates of \kspn\ light curves during the plateau 
phase are $\dot{m}_1$ $\simeq$ 6.58, 3.39 and 2.34 \mhun\ for $B$, $V$, and $i$ band, 
respectively (\autoref{tab:fit_parm}).
The $V$-band decline rate is significantly greater than the mean 
decline rate of 1.27 \mhun\ measured for the sample of Type II SNe in \citet{Anderson_14_sample}.
(Note that $\dot{m}_1$ and $\dot{m}_3$ correspond to $s_2$ and $s_3$, respectively, 
in \citealt{Anderson_14_sample}.)
Large post-peak decline rates during the plateau phase
have been typically observed in SNe with CSM interactions; for example,
in the case of SN~2018hfm, 
which shows near-peak excess emission resulting from 
strong CSM interactions \citep{Zhang_21_Halpha},
an extremely large post-peak decline rate of 4.42 \mhun\ was measured (\autoref{fig:V_LC}). 

The $V$-band rise time of 4.8 days and its short plateau duration of
54.2 days lead to an optically thick phase duration \citep[OTPd;][]{Anderson_14_sample} 
of 59 days for \kspn. 
This OTPd of \kspn\ (\autoref{fig:V_LC}) is  smaller than what 
have been obtained for most of Type II SNe in which
periods larger than 70 days have been measured in most cases
\citep[][see also the example light curve of SN2018gj in \autoref{fig:V_LC}]{Anderson_14_sample}. 
This indicates that \kspn\ likely has a small H envelope 
mass \citep{Popov_93_Henv,Fang_24_Henv}. 

We classify \kspn\ to be a luminous, fast-evolving transitional 
Type II SN with intermediate features between Type II-P and II-L subtypes. 
The presence of its short plateau followed by the drop phase 
classify it to be Type II-P 
\citep[e.g.,][]{Valenti_16_class}, whereas
its decline rate during the plateau phase is 
more than three times greater than those of Type II-L SNe 
\citep[e.g.,][]{Faran_14_class}. 
\kspn\ shares a similar OTPd with the other three 
transitional cases of SN~2006Y, SN~2006ai, and SN~2013by (\autoref{fig:V_LC})
known to have CSM interactions,
but it shows the largest post-peak decline rate among them. 
The near-peak luminosity of \kspn\ is comparable to that of 
SN~2013by \citep{Valenti_16_class},
and it is slightly more luminous than SN~2006Y and SN~2006ai, 
two transitional cases suggested to 
be associated with stripped progenitors \citep{Hiramatsu_21_Halpha}.
(See \S\ref{sec:spec} for our spectroscopic analysis on 
the transitional nature of \kspn.)

\subsection{Colour Evolution}\label{sec:color}

\begin{figure*}
    \centering
    \includegraphics[width=\linewidth]{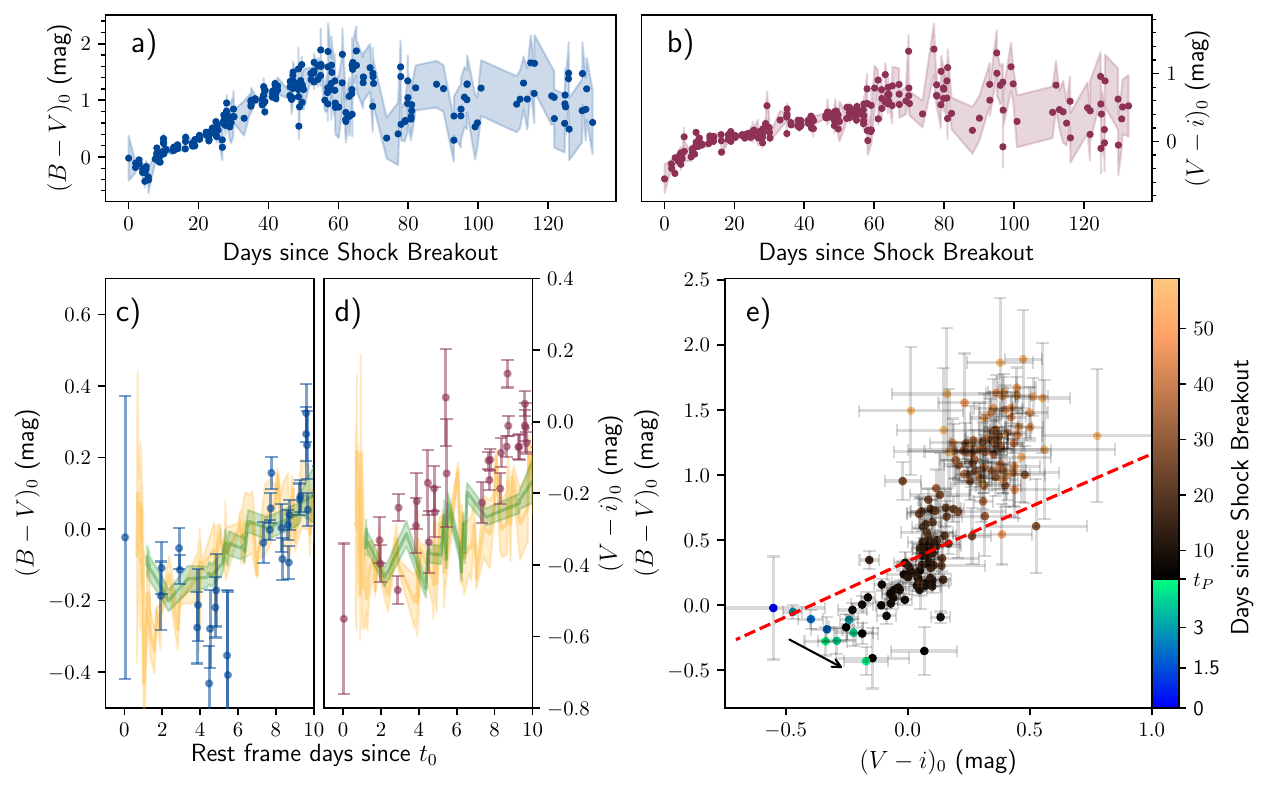}
    \caption{Color evolution of \kspn: (a) and (b) are for \bv\ and \vi\ colours, respectively,
    with the shaded areas representing the uncertainties at the 1$\sigma$ level. 
    (c) and (d) compare the early \bv\ and \vi\ colors, respectively, of \kspn\ 
    (blue and red circles, respectively) with those of SN~2024ggi \citep[shaded yellow;][] {Shrestha_24_ggi} and SN~2023ixf \citep[shaded green;][]{Wang_ifx} in rest frame.
    (e) Colour-colour (\bv\ vs. \vi) diagram of \kspn\
    between $t_0$ and $t_E$ (or $\simlt$ 60 days since SBO) measured in the $V$ band. 
    The colours of the circles represent the time since $t_0$ as shown 
    with the vertical colour bar.
    The black arrow points in the direction of the early colour evolution 
    between $V-$ band $t_0$ to $t_p$ (or $\simlt$ 4 days since SBO) on average.
    The dashed red line shows the colour evolution of a
    blackbody from 30000~K (bottom left corner) to 4000~K (top right corner)
    evaluated at the rest frame isophotal $BVi$ wavelengths.
    All magnitudes are corrected for Milky Way extinctions.}
    \label{fig:colour_comp}
\end{figure*}

\autoref{fig:colour_comp} (a) and (b) present the evolution of 
the \bv\ and \vi\ colours of \kspn\ .
Overall both colours show gradual reddening from SBO
until about 60 days (= $t_E$ in \autoref{tab:fit_parm}) 
after which their evolutions become somewhat flat, 
as often found in Type II SNe \citep[e.g.,][]{Jaeger_19_berkerly}. 
In panels (c) and (d) of \autoref{fig:colour_comp},
we compare the early \bv\ and \vi\ colour evolution of \kspn\ 
with SN~2023ixf and SN~2024ggi,
two CSM interacting Type II SNe
\citep{Wang_ifx, Zimmerman_24_23ixf,Jacobson_24_ggi,Shrestha_24_ggi}.
In the case of \kspn, 
the \bv\ colour becomes bluer by approximately 0.4 mag over the first
$\sim$ 6 days after which it shows reddening by more than $\simgt$ 0.4 mag
during $\sim$ 6--10 days, whereas
the \vi\ colour shows continuous reddening over the first 10 days
by $\sim$ 0.6 mag.
The blueward evolution in \bv\ for SN~2023ixf and SN~2024ggi during the first three days is similar to, but shorter than that of \kspn\ before the colour becomes reddened.
This behaviour of early blueward evolution of the \bv\ colour 
in SN~2023ixf and SN~2024ggi has been interpreted to result from 
shock interactions with the CSM,
either by flash ionization spectral features in $B$
or thermal blackbody heating \citep{Shrestha_24_ggi,Jacobson_24_ggi,Zimmerman_24_23ixf}. 

The early colour evolution of \kspn, 
in which the \bv\ colour shows initial blueward evolution before reddening 
while the \vi\ colour does continuous reddening (\autoref{fig:colour_comp} [c] and [d]), 
is apparently different from what can be expected for simple blackbody heating
as inferred for SN~2023ixf and SN~2024ggi.
As shown in \autoref{fig:colour_comp}(e),
the evolutionary direction of \kspn\ in the \bv\ vs. \vi\ colour-colour plot,
which is denoted by the black arrow, 
is deviated from the expectation (red dashed line) for a blackbody in the temperature
range of 4000--30000 K. 

\subsection{Bolometric Luminosities and \ni56\ Mass} \label{sec:Lum}
\begin{figure*}
    \centering
    \includegraphics[width=\linewidth]{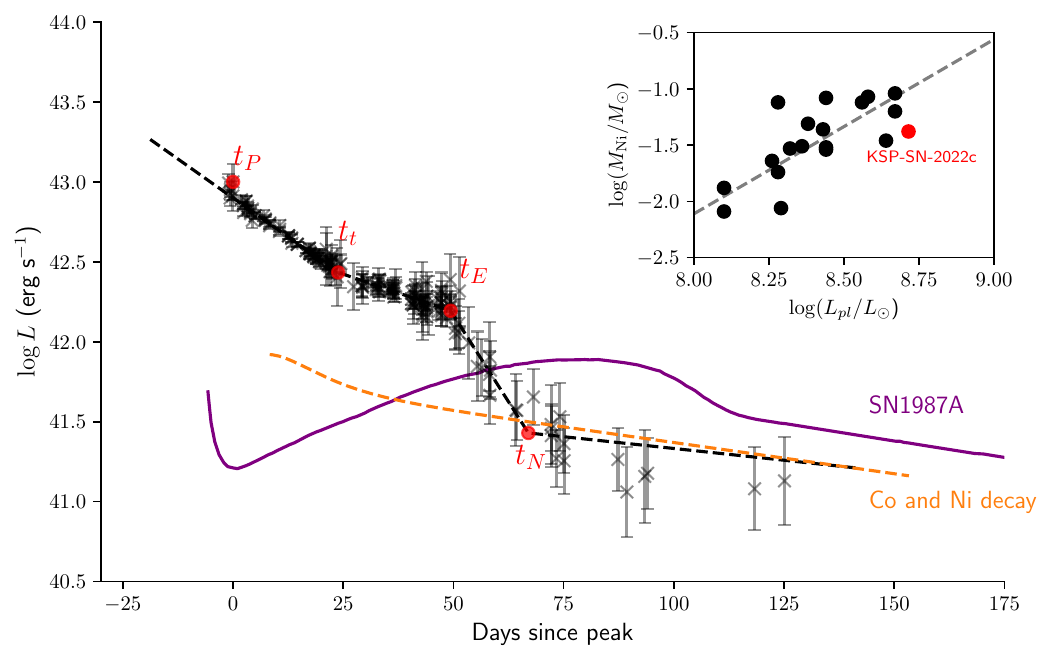}
    \caption{Post-peak bolometric luminosity (crosses with error bars) evolution of \kspn\ 
    overlaid with that of SN~1987A \citep[purple curve;][]{Suntzeff_1987_90}
    and the best-fit luminosity evolution driven 
    by the radioactive decay of \ni56\ and \co56 (dashed orange curve; \S\ref{sec:Lum}). 
    The black dashed lines and filled red circles represent 
    the post-peak decline phases of the bolometric luminosities of \kspn\ 
    determined by \autoref{eq:threephase} (\S\ref{sec:Lum}). 
    The inset compares the plateau luminosities and \ni56\ mass of 17 Type II SNe 
    \citep[filled black circles;][]{Muller_17_Nimass} 
    and \kspn\ (filled red circle) at 50 days since SBO. 
    The dashed grey line shows the best-fit linear relation
    between the two parameters from \citet{Muller_17_Nimass}.}
    \label{fig:bol_comp}
\end{figure*}

\autoref{fig:bol_comp} shows the post-peak bolometric luminosity evolution of \kspn\ 
obtained by applying the \bv\ color correction term from \citet{Lyman_2014_Bolcorr} to its $B$-band light curve. 
The peak bolometric luminosity of \kspn\ is $\simgt$ 1 $\times$ 10$^{43}$ \ergs, 
larger than the range of 10$^{41.5}$--10$^{42.5}$ \ergs\ from most of 
Type II SNe \citep{Martinez_22_bolometric},
making it one of the luminous SNe. 
We identify in the figure the presence of an additional post-peak decline phase
compared to the $BVi$ light curves
(\autoref{fig:raw} and \autoref{eq:threephase}) in the period of 
$\sim$25--50 days since peak. 
Such an additional phase in the post-peak decline
has been interpreted as an indication of the luminosities mainly
driven by H recombination in other Type II SNe
\citep[e.g.,][]{Popov_93_Henv, Martinez_22_bolometric}.
We obtain the starting epoch of this phase 
in the post-peak decline of the bolometric luminosities
to be $t_t$ $\simeq$ 29.6 days since SBO together with 
$t_E$ $\sim$ 55.1 days and $t_N$ $\simeq$ 72.7 days
by fitting  \autoref{eq:threephase} with an additional 
phase for $t_t$ (\autoref{tab:photometry}).

The nebular (or tail) phase luminosity of an SN is known to be 
driven by the radioactive decay of 
\ni56\ $\rightarrow$ $^{56}$Co $\rightarrow$ $^{56}$Fe.
We derive the \ni56\ mass of \kspn\ by applying the following methods 
to its nebular phase bolometric luminosities.
\textit{First}, under the assumption of the complete $\gamma$-ray trapping, the nebular phase luminosities can be modelled as \citep[see e.g.][]{Nilou_21_Nimass}:
\begin{equation}
\begin{split}
L_\gamma=M_{\mathrm{Ni}}\left(\left(\epsilon_{\mathrm{Ni}}-\epsilon_{\mathrm{Co}}\right) \; e^{-t / t_{\mathrm{Ni}}} \; + \; \epsilon_{\mathrm{Co}} \; e^{-t / t_{\mathrm{Co}}}\right)\\
    L_{\mathrm{pos}}=0.034 \; M_{\mathrm{Ni}} \; \epsilon_{\mathrm{Co}}\left(e^{-t / t_{\mathrm{NI}}}-e^{-t / t_{\mathrm{Co}}}\right)
\end{split}
\label{eq:niheating}
\end{equation}
\noindent
where $L_\gamma$ and $L_{\mathrm{pos}}$ denote the luminosities generated from 
$\gamma$-rays and positrons, respectively \citep{Valenti_08_Nimass, Piro_13_Nimass, Nilou_21_Nimass}.
The parameters $\epsilon_{\mathrm{Ni}}=3.9 \times 10^{10}$ \ergs\
and $\epsilon_{\mathrm{Co}}=6.8 \times 10^9$ \ergs\ are 
the specific heating rates for \ni56\ and $^{56}$Co, respectively, while
$t_{\mathrm{Ni}}=8.8$ days and $t_{\mathrm{Co}}=111.3$ days are the decay timescale for \ni56\ and $^{56}$Co.
By fitting the equation to the bolometric luminosities of \kspn\ after $t_N$ (orange curve), 
we obtain \mni56\ = $0.042$ $\pm$ 0.003 \msol.

\textit{Secondly}, we compare the fitted nebular-phase bolometric luminosities 
of \kspn\ (\autoref{fig:bol_comp}) to the bolometric luminosities of SN~1987A 
from \citet[purple curve][]{Suntzeff_1987_90}
during the 115--130 days period since peak.
As a result, we obtain  \mni56 $\sim$ 0.043 \msol\ for \kspn,
similar to what we have obtained above,
using  0.075 \msol\ for the \ni56\ mass of SN~1987A \citep{Arnett_Supernovae_1996}.
We adopt \mni56\ = $0.042$ $\pm$ 0.002 \msol\ from the assumption
of the complete $\gamma$-ray trapping 
as the \ni56\ mass of \kspn, and this follows 
the known relation between \ni56\ mass and plateau 
luminosities at 50 days since SBO of Type II SNe 
from \citet[][see also the top panel in \autoref{fig:bol_comp}]{Muller_17_Nimass}.
The \ni56\ mass of \kspn\ is slightly larger than the average ($\simeq$ 0.032 \msol) 
value of Type II population \citep{Anderson_19_Nimass}. 

\section{Spectroscopic Analysis and Interactions}\label{sec:spec}

\begin{figure}
    \centering
    \includegraphics[width=\linewidth]{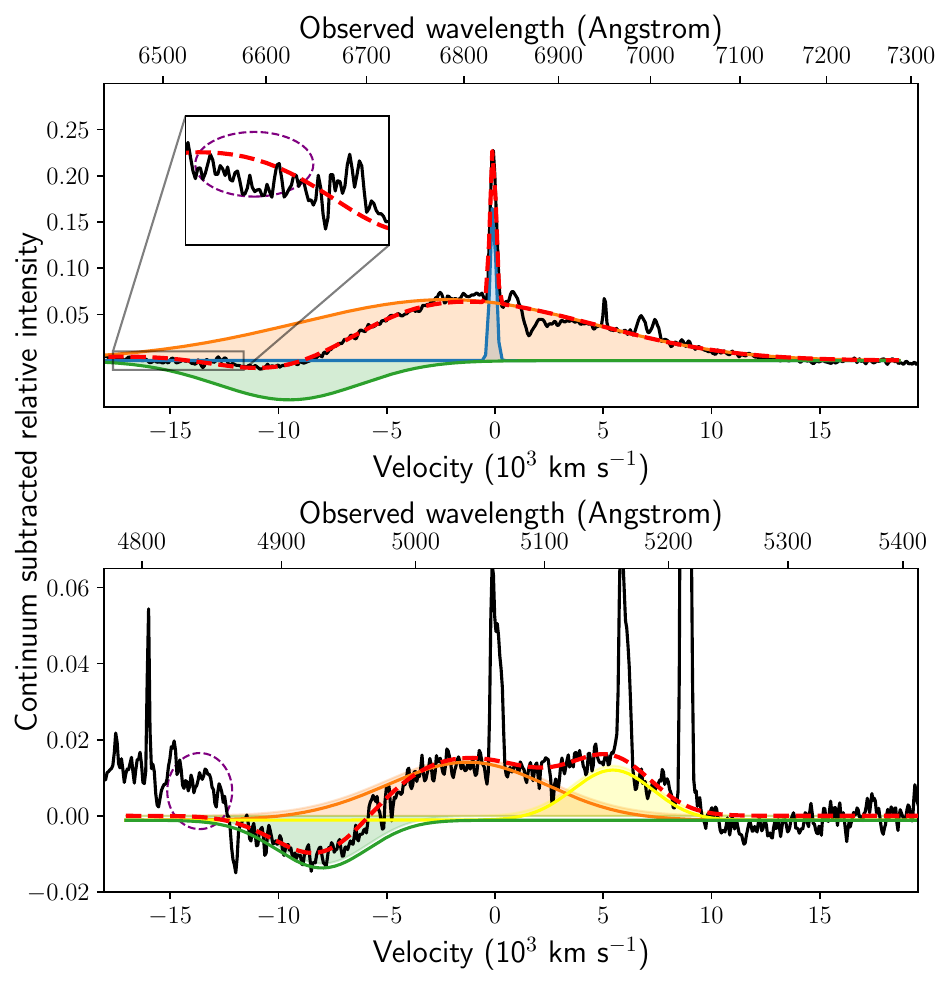}
    \caption{({\it Top Panel}) The observed \hal\ spectrum (black curve) of \kspn\ 
    and fitted Gaussian components:
    the blue curve for narrow emission, the orange curve for broad emission,
    and green curve for broad absorption (\autoref{table:measurements2}). 
    A flat baseline for the continuum is fitted together and subtracted.
    The dashed red curve represents the summation of all the 
    fitted components.
    The dashed ellipse in the inset highlights the discrepancy 
    between the observations and fitted components likely caused
    by the presence of a high-velocity absorption component (see text).
    ({\it Bottom Panel}) Same as the top panel, but for \hbe\ with
    the broad emission (orange curve) and broad absorption (green curve) components,
    together with Fe II ($\lambda$4924) line emission (yellow curve). 
    The two narrow peaks between +5000 and +10000 \kms\ are from  
    the \oiii\ doublet ($\lambda\lambda$4963, 5006).}
    \label{fig:spec_fea}
\end{figure}

\autoref{fig:spec_fea} show our results of three component Gaussian fitting
of the observed spectrum of \kspn\ in \autoref{fig:spec_all} for the
\hal\ (top panel) and \hbe\ (bottom panel) features.
The fitted three Gaussian components of \hal\ are for
broad emission component (red curve),
blue-shifted absorption component (green curve), 
and the narrow peak (blue curve).
For \hbe, the three Gaussian components are
for broad emission component (red curve), 
blue-shifted absorption component (green curve), 
and the broad nearby \feiio\ ($\lambda$ 4924) emission (yellow curve).
In the fitting procedure, we exclude
the narrow \hbe\ peak at zero velocity as well as
two narrow peaks for \oiii\ doublet at 6000 and 9000 \kms\
from the host.
The dashed red curves in the figure represent the summations of the 
three fitted Gaussian components for \hal\ (top) and \hbe\ (bottom).
\autoref{table:measurements2} lists the best-fit parameters of the 
fitted Gaussian components. 

\begin{table*}[]
\centering
\caption{Best-fit Gaussian parameters of the components in the \hal\ and \hbe\ features}
\begin{tabular}{lcc}
\hline
\hline
Components & Central Wavelength (\AA) & FWHM (\AA) \\
\hline
Narrow \hal\ emission & 6829.13 $\pm$ 0.08 & 7.05 $\pm$ 0.20 \\
Broad \hal\ emission & 6782.88 $\pm$ 6.61 & 360.27 $\pm$ 7.65 \\
Broad \hal\ absorption & 6626.58 $\pm$ 2.08 & 165.48 $\pm$ 8.71 \\
\hline
Broad \hbe\ emission & 5040.49 $\pm$ 6.26 & 145.15 $\pm$ 43.11 \\
Broad \hbe\ absorption & 4934.02 $\pm$ 4.63 & 79.37 $\pm$ 9.72 \\
Broad \feiio\ emission & 5157.90 $\pm$ 4.31 & 76.32 $\pm$ 11.16 \\
\hline
Note: The wavelengths are observed values.
\end{tabular}
\label{table:measurements2}
\end{table*}

We measure the FWHM of the broad \hal, \hbe, and \feiio\ ($\lambda$5157) 
emission directly from the observed spectrum of \kspn\ 
to be $\sim$ 260, 120 and 73 \AA, respectively, in rest frame. 
These values of FWHM of emission components that are partly overlapping 
with absorption (\autoref{fig:spec_fea})
translate to ejecta expansion velocities of 
$\sim$ 11000, 6800 and 4400 \kms, respectively,
according to \citep{Guti_14_Halpha}.
If we use the fitted central wavelengths of the broad 
absorption components in \hal\ and \hbe\ in \autoref{table:measurements2},
the corresponding ejecta velocities are $\sim$ 9700 and 8000 \kms, respectively. 
The expansion velocities of \kspn\ from the two methods are largely comparable,
while they are somewhat larger than the average values obtained for a sample of Type II SNe at similar epochs \citep{Gutirrez_17_Specsample}:
6000$\pm$ 1500 and 6500 $\pm$ 1800 \kms\ from absorption and emission 
measurement, respectively, for \hal\ and 
5000 $\pm$ 1800 \kms\ for \hbe\ .

We also measure the \hal\ absorption-to-emission equivalent width ratio ($a/e$) 
of \kspn\ from the observed (not fitted) spectrum to be $a/e$ $\simeq$ 0.076 
following the definition in \citet{Guti_14_Halpha}.
This is one of the smallest $a/e$ ratios among Type II SNe,
while it is similar to the values obtained for CSM-interacting events
such as SN~2006Y (\autoref{fig:spec_all}), SN~2006ai, and SN~2018hfm
\citep{Hiramatsu_21_Halpha, Zhang_21_Halpha}. 
Small $a/e$ ratios have been interpreted to result from the suppression
of absorption in P-Cygni profiles by CSM interactions \citep[e.g.,][]{Guti_14_Halpha},
and are usually associated with events with large peak luminosities, 
fast post-peak declines, as well as large ejecta velocities
\citep{Guti_14_Halpha, Gutirrez_17_Specsample, Dessart_22_CSM} 
as in \kspn. 

\autoref{fig:spec_fea} shows the potential presence of  high-velocity ($\sim$ --14000 \kms),
blue-shifted features near the absorption end of the  \hal\ and \hbe\ profiles
with the dashed ellipses in the figure. 
Similar high-velocity absorption features,
which have been attributed to strong reverse shock interactions in CSM \citep{Chugai_07_CSM}, 
have been identified in a few other Type II SNe---including 
SN~2018hfm \citep{Zhang_21_Halpha}, 2018lab \citep{Pearson_23_2018lab}, 
and 2023axu \citep{Shrestha_24_2023axu}, and 
as well as a sample studied by \citet{Gutirrez_17_Specsample}---showing
signature for CSM interactions.

We identify in \autoref{fig:spec_all} the observed spectrum of \kspn\ 
is overall very similar to those of 
SN~2006Y  and 2016egz---two luminous, fast-declining Type II SNe 
with a short plateau and CSM interactions \citep{Hiramatsu_21_physical},
while it is different from the three 
(SN~2013ab, SN~2015V, and SN~2016X) normal Type II-P SNe 
obtained at similar epochs.

\section{Light Curve Modeling}
\label{sec:modeling}
\subsection{Semi-Analytic Models for the Early Light Curve: Shock Cooling Emission and CSM Interaction}

The early emission from CCSNe is dominated by shock cooling emission (SCE)
following SBO. The SBO occurs when the shock velocity is $\varv_s$ $\simeq$ $c/\tau$, where $\tau$ is the optical depth \citep{Ohyama_63_SBO},
at which light can escape from the ejecta.
\citet{SMW23_SCE} recently developed an SCE model combining both 
planar and spherical shock phases based on the previous work by 
\citet{KSW12} and  \citet[][see \citealt{RW_11_SCE} for original work]{SW17_SCE}, respectively.
This model is applicable during early phases before the decrease in opacity 
caused by recombination becomes significant (at $T\simlt 0.7$ eV).
The evolution of temperature and luminosity of the model is determined by 
shock velocity, ejecta mass, and other parameters for stellar structure. 

\begin{figure*}
    \centering
    \includegraphics[width=\linewidth]{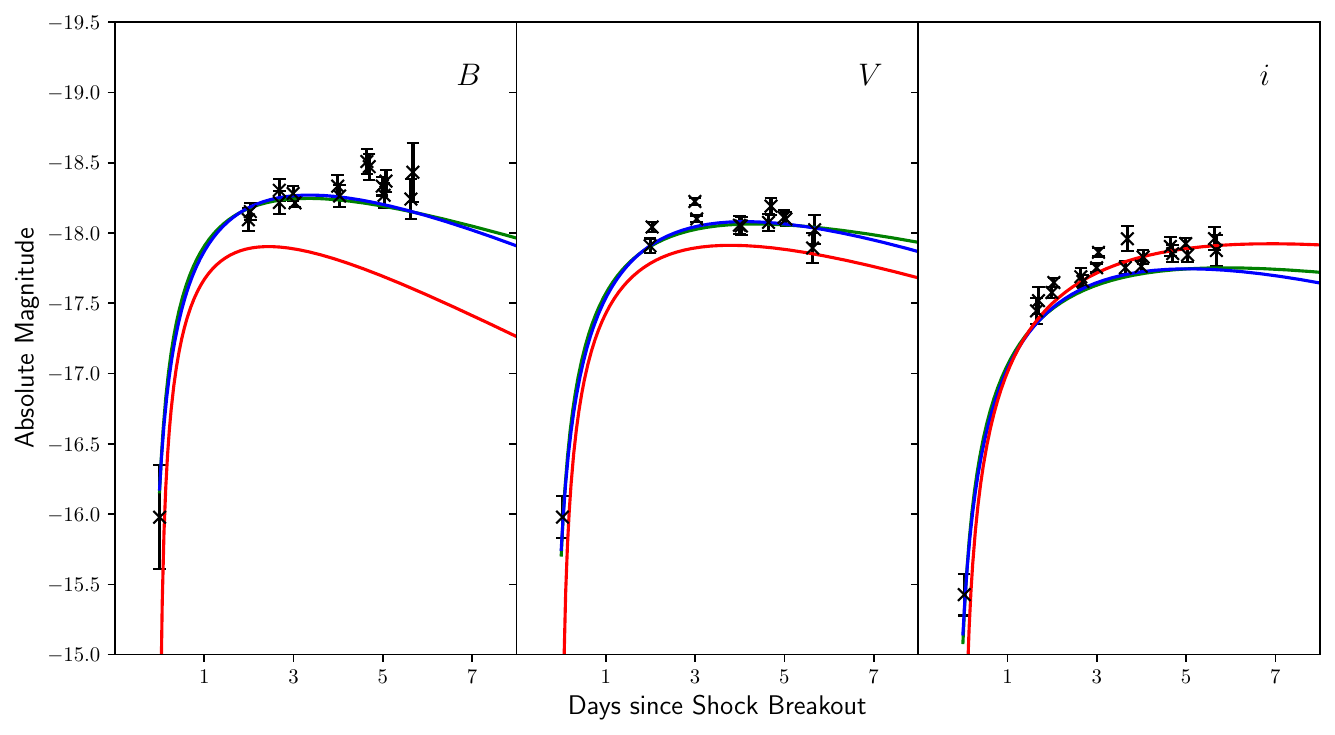}
    \caption{Comparisons between the observed $BVi$ light curves 
    (crosses with error bars for one sigma uncertainties) 
    of \kspn\ and the best-fit predictions from two SN SCE models 
    of \citet[][green]{SMW23_SCE} and \citet[][blue]{SW17_SCE} 
    obtained by fitting the models to the observed light curves 
    before the peak.
    The red curves represent the best-fit predictions by 
    \citet[][]{Piro21} for emission from CSM interacting with SN shocks.}
    \label{fig:LC_fit}
\end{figure*}

We apply the SCE model by \citet{SMW23_SCE} to \kspn\ before the $B$-band 
light curve peak at 4.5 days from SBO assuming blackbody radiation for the
observed $BVi$ light curves.\footnote{We use the Python package
\texttt{Light Curve Fitting} \citep{LC_fit} to calculate the model predictions.}.
\autoref{fig:LC_fit} (green curves) shows the best-fit case obtained with 
$\varv_{s} =\left(2.9_{-0.3}^{+0.6}\right) \times 10^{8.5} \mathrm{~cm} \mathrm{~s}^{-1}$, 
envelope mass $ M_{\text {env }}=0.22_{-0.04}^{+0.06}$ \msol, 
and progenitor radius $R=\left(4.0_{-0.9}^{+0.6}\right) \times 10^{13} \mathrm{~cm}$, 
As in the figure, while the best-fit case 
matches the $V$-band light curve reasonably well, it 
underpredicts $B$- and $i$-band fluxes around the peak,
especially in the $B$ band, indicating the presence of additional emission. 
The best-fit prediction for SCE by the spherical shock model 
\citep[][blue curves in \autoref{fig:LC_fit}]{SW17_SCE} is
almost identical to that by the model of \citet{SMW23_SCE} (green curves)
except that the former produces slightly lower luminosities at 
later ($\simgt$ 7 days) epochs, still under-predicting in the $B$ and $i$ bands.

Interactions between the CSM and ejecta can contribute
significantly to early SN luminosities \citep[e.g.,][]{yao_20_doublepeak, Pellegrino_23_doublepeak}.  
\citet{Piro21} provides the evolution of the photospheric luminosity and radius from shock interactions between 
SN ejecta and CSM by adopting two power-law forms for 
the CSM density profile \citep{Chevalier_soker_89}, 
leading to a two-zone evolution for them. 
We fit the model predictions by \citet{Piro21}
under the assumption of blackbody radiation
to the observed $BVi$ light curves of \kspn. 
As in \autoref{fig:LC_fit} (red curves),
we can identify that the ejecta and CSM interaction model
by \citep{Piro21} does not match the observed light 
curves with significant
under-predictions for the $B$ and $V$ bands,
strongly indicating that the CSM interactions alone 
are insufficient to explain the observed light 
curves  of \kspn. 
Below, we describe our numerical simulations of 
\kspn\ light curves by taking both luminosities 
from an SN explosion and CSM interactions into account.

\subsection{Modeling of Light Curves with SNEC Simulations
}\label{sec:SNEC_model}
Using the SuperNova Explosion Code
\citep[SNEC;][]{SNEC_15, SNEC_16_early,SNEC_17_CSM, SNEC_18_CSM}, 
which is based on 1D Lagrangian radiation hydrodynamics,
we conduct light curve simulations of SN explosions 
to constrain the physical parameters required to 
produce observed light curves of \kspn.
For all our simulations with SNEC, we closely
follow the steps adopted in \citet{SNEC_17_CSM,SNEC_18_CSM},
including the following
five: (1) fixing the neutron star mass to be 1.4 \msol;
(2) injecting the explosion energy into the innermost 
0.02 \msol\ of the progenitor mass for a duration of one second as a thermal bomb;
(3) adopting the equation of state from \citet{Paczynski83} 
that takes into account contributions from ions, electrons, and radiation;
(4) limiting the comparisons between the observed and simulated
light curves only to the end of the plateau phase after which
the assumptions of LTE and black body radiation in SNEC
become no more valid; and
(5) excluding the $B$-band light curves from comparisons 
due to the substantial contributions from line emission 
in that band.

In our simulations, we use SN progenitor profiles 
from the \texttt{KEPLER} stellar evolution code 
\citep{Kepler_98, Kepler_07, Kepler_14, Kepler_15, Kepler_16}
for evolved non-rotating red supergiant (RSG) models with solar metallicity.
We also use 0.042 \msol\ for the amount of $^{56}$Ni of 
\kspn\ (\S\ref{sec:Lum}) in two mixing schemes of up to 3 
and 5 \msol\ in the mass coordinate. 

\subsubsection{Baseline Models: Explosion of RSG Stars} \label{sec:baselinemodel}

For our initial baseline model fits, 
which include only expectations for RSG stars with 
structures described by the \texttt{KEPLER} models, 
we fit only the data in the interval $\simeq$30-60 days. 
This excludes the region of more rapid decline observed 
in the bolometric light curve (see \autoref{fig:bol_comp})
which has been interpreted to be the period with contributions
from CSM interactions in other SNe \citep[e.g.,][]{SNEC_18_CSM},
and similar exclusions have been made in previous studies \citep[e.g.,][]{SNEC_17_CSM, SNEC_18_CSM}. 
(Effects of CSM interactions will be described in more detail below.)

The best-fit light curves for \kspn\ are produced 
by the combination of a progenitor with 13.0 \msol ZAMS mass
and 680 \rsol\ radius 
and 1.05 foe explosion energy.
The grids of our simulations are 
9--20 \msol\ with a 0.5 \msol\ increment for progenitor mass
and  0.5--2.0 foe with an increment of 0.05 foe for explosion energy, similar to studies such as \citet{SNEC_18_CSM} and \citet{Martinez_22_gridsample}.
As we can identify in \autoref{fig:sim_big} (left panel),
the best-fit $Vi$-band light curves (dashed curves)
show a reasonable agreement to the observed light curves of \kspn\
(crosses with error bars) in 30--60 days where comparisons are made.
However, their extended light curves significantly under-predict luminosities compared to the observed ones during earlier epochs of $<$ 30 days, while they fail to produce the observed rapid decline 
at later epochs of $>$ 60 days.

\subsubsection{Best Fitted Baseline Model Plus CSM Interactions: Impact on Early Luminosity}

The early excess emission in the light curves of CCSNe,
as observed in \kspn\ (\autoref{fig:LC_fit}),
has often been attributed to the results of CSM interactions. To examine whether CSM interactions can indeed be responsible for the excess emission in \kspn\, 
we conduct SNEC simulations with CSM-added progenitor profiles. We attach CSM density profiles (\autoref{fig:sim_big}, top-right panel) 
taking a form of $\rho_{\rm CSM} (r)$ = $K_{\rm CSM}$/$r^2$,
where the mass loading factor $K_{\rm CSM}$ is determined by the mass 
loss ($\dot{M}$) and wind terminal velocity ($\varv_{\infty}$) as $K_{\rm CSM}$ = $\dot{M}$/4$\pi$$\varv_{\infty}$ \citep{Parker_wind},
to the progenitor profiles.
In this case,  the total mass of the enclosed CSM is $\Mcsm$ = 4$\pi\Kcsm\Rcsm$ 
where $\Rcsm$ is the radius of the CSM measured from the progenitor radius. $\Rcsm$ and $\Kcsm$ have been measured in the ranges of
700--3000 \rsol\ and (1 $\times$ $10^{17}$)--(1 $\times$ $10^{18}$) g~cm$^{-1}$, respectively,
in CCSNe \citep{SNEC_17_CSM, SNEC_18_CSM, Haynie_21_CSM, Nilou_21_Nimass, meza_2021gmj_24}. 
Therefore, in our simulations, we vary $\Kcsm$ in the ranges of 
(1 $\times$ $10^{17}$)--(3 $\times$ $10^{18}$) g~cm$^{-1}$ 
with an increment of 1 $\times 10^{17}$ g~cm$^{-1}$ and $\Rcsm$  in the range of
700--3800 \rsol\ with an increment of 100 \rsol.

We then attach the CSM density profile constructed based on different combinations of $\Kcsm$ and $\Rcsm$ 
to the best-fit RSG progenitor profile (\ref{sec:baselinemodel}).
\autoref{fig:sim_big} (solid curves) shows the best-fit CSM-added simulations to the observed $V,i$-band light curves of \kspn\ 
during the period of 0--60 days since SBO 
obtained by the combination of  
\rcsm\ $\simeq$ 1900 \rsol\ (= 2.8 times of the progenitor radius) and
\kcsm\ $\simeq$ 9 $\times$ 10$^{17}$ g~cm$^{-1}$.
In contrast to the progenitor-alone fit (dashed curves),
the CSM-added fit matches the observed light curves,
including the early excess emission,
during the period of 0--60 days even for the $B$ band which was not included in the fits.
At epochs $>$ 60 days, which are not included in the fitting, 
the observed light curves decline much faster with a shorter
plateau than the best-fit simulation.
This could be because \kspn\ becomes
optically thin at these epochs where SNEC is no longer applicable
(\S\ref{sec:baselinemodel}). 
(We explore in \S\ref{sec:stripping} a possibility that
the observed short plateau of \kspn\ is a consequence of significant stripping of the progenitor's envelope.)

Under the assumption a steady wind, the best-fit values of \rcsm\ and \kcsm\ lead to the 
CSM mass of 0.73 \msol. 
This is greater than most of the previously estimated CSM mass of Type II SNe \citep[e.g.,][]{SNEC_18_CSM}, 
while it is comparable to the case of SN~2013by,
a transitional type-II SN between II-P and II-L \citep{SNEC_17_CSM}. 
We note that it is difficult to constrain \rcsm\ and \kcsm\ 
simultaneously because their product ultimately determines CSM contributions
to light curves (\autoref{fig:sim_big} bottom right panel).
The SBO of the best-fit case of our CSM-added simulations 
occurs at a radius of about 3.7 times the progenitor radius,
which is almost the same as the CSM radius measured from the progenitor centre, suggesting that the location of SBO is at the outer edge 
of the CSM.

The best-fit light curves of our CSM-added simulations (\autoref{fig:sim_big}, solid curves in the left pane) 
begin with almost vertical rises of 
$\simlt$ 2 mags at $\simlt$ 20 minutes from SBO, 
which have also been seen in other
CSM-added SNEC simulations using steady wind \citep[e.g.,][]{SNEC_18_CSM, Dastidar_25_SNECflaw}.
Considering the absence of such a feature in
similar SNEC simulations with non-steady wind CSM
configuration \citep[see][for non-steady wind case]{Moriya_18_CSMaccerlation, Forster_18_earlyCSM},
the feature is likely caused by the use of a steady 
wind for the CSM density profile. 
Since this feature occurs before our first detection
of \kspn\ in all the three $BVi$ bands and not
included in our fitting,
it does not affect the results of our simulations. 

\subsubsection{Best Fitted Baseline Model with Envelope Stripping: Impact on Plateau Duration}\label{sec:stripping}

The main difference between the observed light curves of \kspn\
and aforementioned our SNEC simulations are the presence of 
the observed short plateau.
One potential way to explain this is partial progenitor envelope stripping 
before the SN explosion. 
The duration and brightness of plateaus of Type II SNe 
are  known to be directly related to 
the envelope mass \citep[e.g.,][]{Popov_93_Henv, Kasen_09_Henv, Kepler_16}.
According to recent simulation studies 
by \citet{Fang_24_Henv},
the plateau duration is mainly determined by three 
parameters---progenitor radius, envelope mass, 
and explosion energy---with envelope mass being the
most decisive parameter (see also \citealt{Popov_93_Henv}).
The plateau duration increases along the envelope mass,
but decreases as the progenitor radius or explosion energy increases. 

\begin{figure*}
    \centering
    \includegraphics[width=\linewidth]{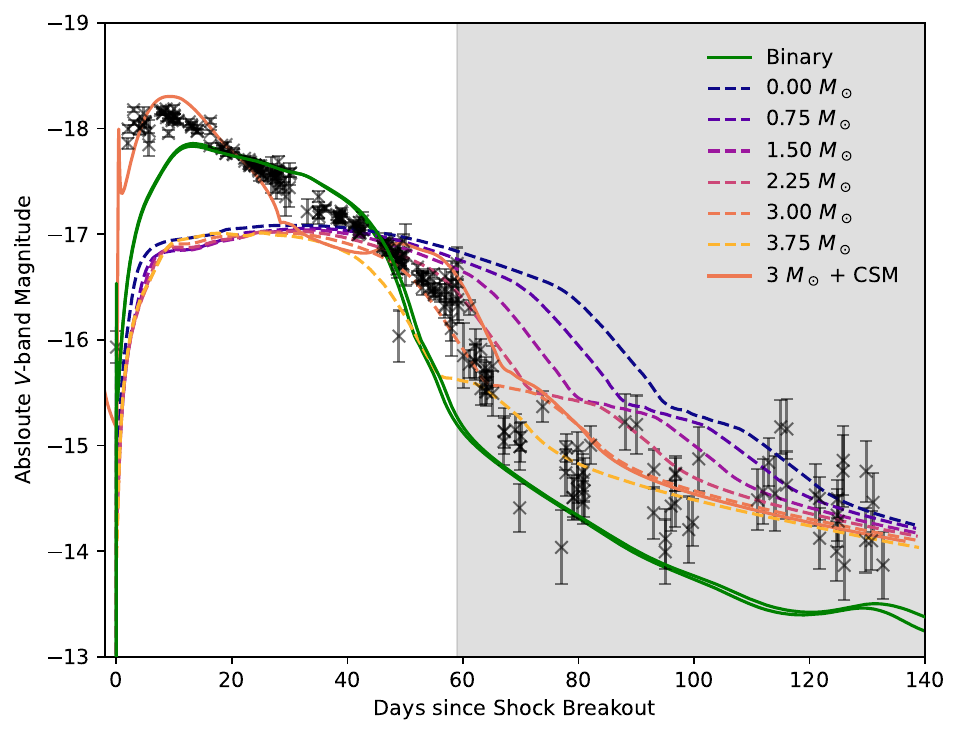}
    \caption{Comparison between the observed light curve of \kspn\ 
    (black crosses with error bars) with six SNEC-simulated $V$-band light curves (coloured dashed curves) of CCSNe 
    from a progenitor with 13 \msol\ and 1.05 foe of energy.
    The simulated light curves are for envelope stripping  of 0.75 (dark purple), 
    1.50 (light purple), 2.25 (cabaret). 3.0 (red), and 3.75 (yellow) \msol\ 
    as well as no envelope stripping (blue).
    No CSM component is included in these six simulations.
    The solid orange curve is the case of 3.0 \msol\ envelope stripping 
    but with a CSM component of 0.73 \msol\ that provides the best match to the observed plateau.
    The solid green curves are the two closest matches to the observed light curve 
    out of 158 SNEC-simulated SN light curves of binary progenitors \citep{Eldridge_18_binary}.
    The primary mass and initial binary period for the two cases are 12 \msol\ and 1000 days, respectively,
    but they have different mass ratios of 0.2 or 0.3.}
    \label{fig:strip}
\end{figure*}

In order to explore the effects of the envelope 
stripping to the shape of SN light curves, 
especially the duration of the plateau phase, 
We first run CSM-free SNEC simulations using the best-fit progenitor (\S\ref{sec:baselinemodel}), where the outermost envelope is removed by masses of 0.75, 1.5, 2.25, 3, and 3.75 \msol, while maintaining the shape of the density profile.
\autoref{fig:strip} shows the $V$-band light curves
(coloured dashed curves) obtained by the SNEC simulations with
stripped-envelope progenitor of 13 \msol\ and 1.05 foe of energy
for the different mass stripping.
We can identify in the figure that shorter plateaus are produced when more mass is stripped,
with the case of 3 \msol\ stripping being the best match to the observed plateau of \kspn.
Note that the early luminosities are under-predicted
due to the absence of CSM. 

\begin{figure*}
    \centering
    \includegraphics[width=\linewidth]{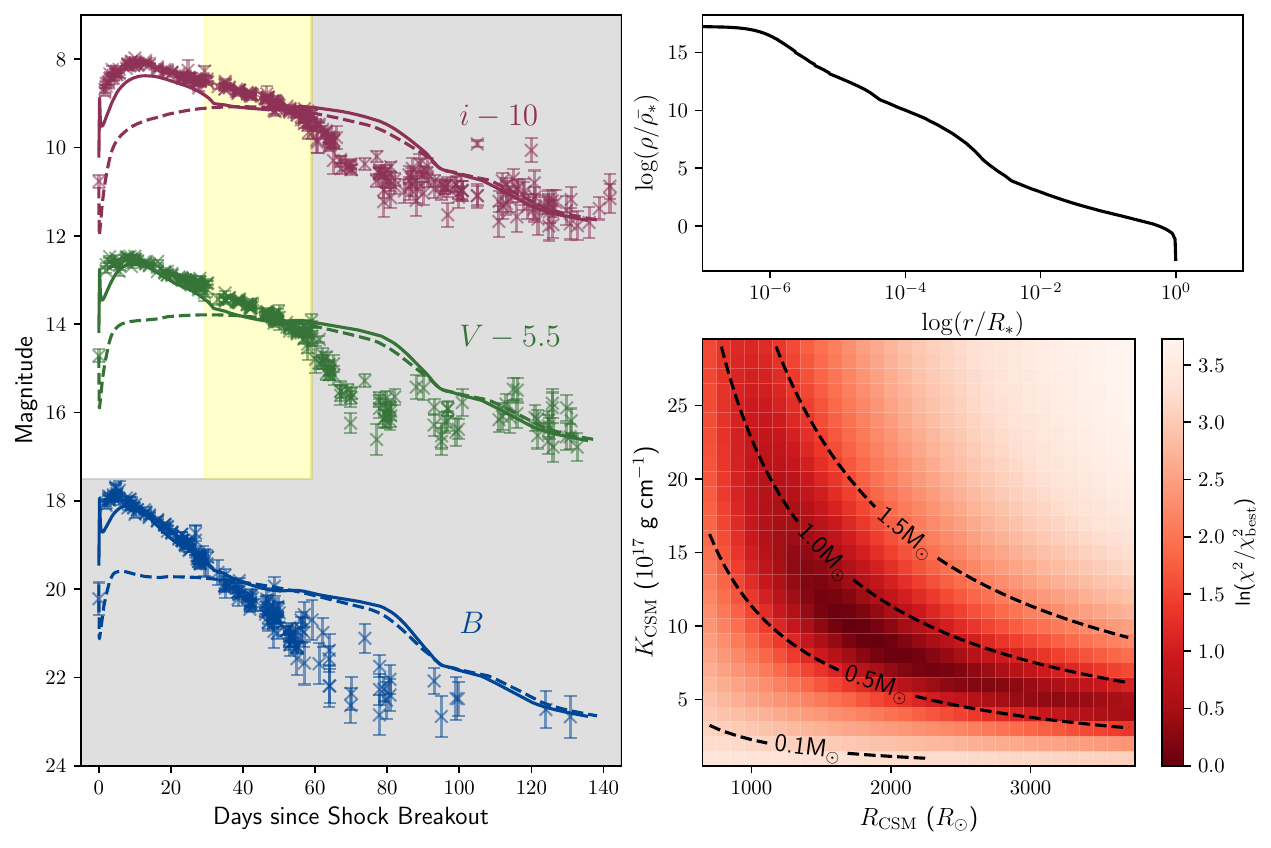}
    \caption{({\it Left Panel}) Comparisons between the observed $BVi$ light curves 
    (crosses with error bars) of \kspn\ and the best-fit 
    SN simulations using stellar profiles for progenitor alone 
    (dashed curves) and progenitor plus CSM components (solid curves).
    The $V$- and $i$-band light curves are shifted vertically to avoid 
    overlapping by 5.5 and 10 magnitudes, respectively. 
    The light curves on the grey-shaded area are excluded 
    from the fit either because the data points
    are obtained after 60 days since SBO or they are in the $B$ band.
    The light curves on the yellow shaded area are used in the progenitor alone fit, 
    while those in the white-shaded area are used in the progenitor plus CSM fit.
    ({\it Top-right Panel}) The density profile of the best-fit
    progenitor from the centre (left) to the outer edge (right)
    scaled by its radius (abscissa) and mean density (ordinate).
    ({\it Bottom-right Panel}) 
    Contours of the CSM mass $M_{\rm CSM}$ = 0.1, 0.5, 1.0, 
    and 1.5 \msol\ as a function of \rcsm\ and \kcsm\ 
    from the progenitor plus CSM fit. 
    The darker colours represent areas with smaller $\chi^2$ values
    as shown on the colour bar.}
    \label{fig:sim_big}
\end{figure*}

The solid red curve in \autoref{fig:strip} shows the $V$-band light curve 
obtained by SNEC simulations with 0.73 \msol\ CSM (\autoref{fig:sim_big} solid curves) 
and 3 \msol\ envelope stripping.
The light curve with the CSM and envelope stripping 
reproduces the observed key features of \kspn\ well,
including large early luminosities, 
rapid post-peak decline, and short plateau.
This is indicative that \kspn\ light curve is a result 
of an SN explosion of a progenitor with substantial CSM 
and envelope stripping. 

\section{Discussion}\label{sec:discuss}

\kspn\ is an infant, luminous ($M_V$ $\simeq$ --18.41 mag at peak) Type II SN 
from a barred spiral host galaxy at
$z$ $\simeq$ 0.041
showing fast rises and rapid post-peak declines with a short plateau of OTPd $\simeq$ 60 days.
We find excess emission near the peak to what is expected from SN shock cooling emission.
There exists a variety of observed features supporting the presence of CSM interactions
for this SN, including its large peak luminosity,
early colour evolution,
broad and asymmetric \hal\ feature with a small $a/e$ ratio 
and high-velocity components.
Our numerical simulations of SN light curves 
show that a progenitor with 13 \msol\ and 1.05 foe for the progenitor mass and SN explosions energy, respectively, 
is the best match to the observed light curves of \kspn. 
It's near-peak excess emission and short plateau duration
are attributable to interactions between the SN and 
CSM of $\sim$ 0.73 \msol\ and progenitor envelope stripping
of $\sim$ 3 \msol, respectively.

Below we discuss the transitional nature of \kspn\ 
and establish a correlation between CSM mass and post-peak decline rate among
Type II SNe.
We also provide our interpretation of the early colour evolution
of \kspn\ in the context of a delayed SBO near the outer edge of CSM,
and explore the possibility of binary progenitor for the source 
more suitable for substantial mass loss.

\subsection{CSM Interactions and Transitional Nature of \kspn}\label{sec:dis_tran}

\subsubsection{Transitional Nature}

\kspn\ shows the presence of a plateau (\autoref{fig:V_LC}) as found in other Type II-P subtypes of CCSNe, 
and the II-P nature of the source is also confirmed by its post-plateau 
drop phase (\S\ref{sec:lc}).
The plateau of \kspn, however, is very short with a rapid decline rate 
$s50_V$ $\simeq$ 1.7 mags per 50 days,
which is much faster than those of the majority of the Type II-P population while similar to the Type II-L population.
The $s50_V$ $\simeq$ 1.7 mags per 50 days
is also higher than 0.5  mags per 50 days, 
which is suggested as the dividing criterion between II-P and II-L subtypes \citep{Faran_14_class}.
This transitional nature of \kspn\ exhibiting characteristics of both II-P and II-L subtypes
support the interpretation that the two subtypes are not intrinsically different 
but rather  represent a continuous distribution \citep[e.g.,][]{Faran_14_class, Faran_14_population, Anderson_14_sample, Sanders_15_population}. 

\begin{figure}
    \centering\includegraphics[width=\linewidth]{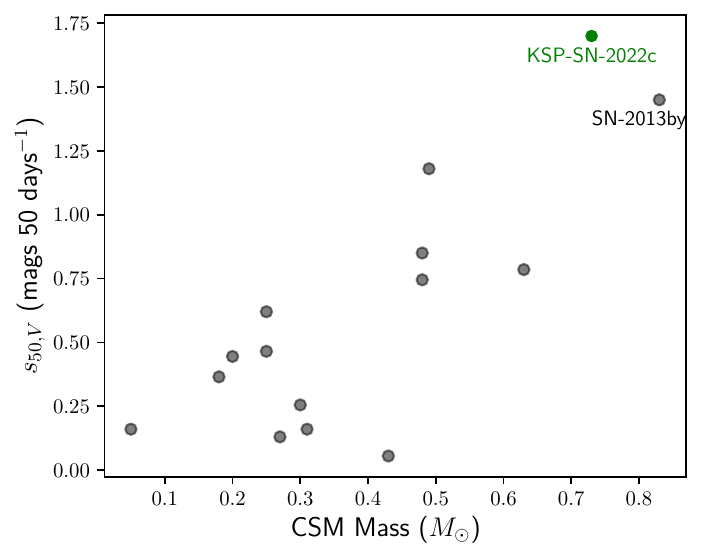}
    \caption{Distribution of CSM mass and \sfif\ parameters of  15 Type II SNe: 
    14 from \citet{Valenti_16_Sndiversty, SNEC_18_CSM} and \kspn\ from this study.
    The two transitional events of SN~2013by and \kspn\ have the largest CSM mass 
    and \sfif\ parameters.}
\label{fig:csms50}
\end{figure}

\autoref{fig:csms50} compares the CSM masses and \sfif\ parameters 
of 15 Type II SNe: 13 Type II-P \citep{SNEC_18_CSM}, 
one transitional  \citep[SN~2013by;][]{Valenti_16_Sndiversty}, and \kspn. 
We identify a clear correlation between the two parameters  when CSM mass 
is determined by the same method described in \S\ref{sec:SNEC_model} 
with the $p$ parameter of the Spearman rank-order test smaller than 0.001.
Notably the two transitional cases of \kspn\ and SN~2013by 
show the largest CSM mass of 0.73 (\kspn) and 0.83 \msol\ (SN~2013by) 
with the fastest post-peak decline rates when compared with $s50_V$ parameters.
This strongly indicates that within the model framework, the amount of CSM is critical to the 
post-peak decline rates of Type II SNe, with more CSM mass 
most likely leading to more rapid declines,
and that transitional events are those with substantial CSM
interactions as previously suggested
\citep[e.g.,][]{,Valenti_16_class,SNEC_18_CSM,Hiramatsu_21_physical}.

\subsubsection{Early Multi-Color Evolution and Delayed Breakout}

The early \bv\ colour evolution of \kspn\ within the first few days from SBO 
shows short-lived blueward evolution followed by reddening after $\sim$ 6 days (\autoref{fig:colour_comp}).
Early short-lived blueward evolution in \bv\ has also been reported 
in two recent CSM-interaction Type II SNe: SN~2023ixf \citep{Wang_ifx, Hosseinzadeh_23_ixf, Zimmerman_24_23ixf} and SN~2024ggi \citep{Shrestha_24_ggi, Jacobson_24_ggi}.
In contrast to the \bv\ colour evolution, 
the  \vi\ colour evolution of \kspn\ shows continuous reddening
during this early phase without a sign of such blueward evolution.

One possible explanation for the observed early multi-colour evolution of \kspn\ 
in \bv\ and \vi\ 
is a delayed SBO near the outer edge of a massive CSM.
It has been suggested that SBO from a progenitor with massive CSM
can be delayed \citep[see, e.g.,][]{Forster_18_earlyCSM, Shrestha_24_ggi, Zimmerman_24_23ixf}.
If a delayed breakout occurs near the outer edge of massive CSM leading to  
a small fraction of unshocked material at the moment of SBO,
it can result in less substantial thermal heating.
As described in \S\ref{sec:modeling}, our numerical simulations of the 
observed light curves of \kspn\ require the presence of 
dense, massive CSM of $\sim$ 0.73 \msol\
as well as a delayed SBO at a distance of 3.7 times the progenitor radius 
near the outer edge of the CSM.
In this scenario,
the early short-lived blueward evolution in \bv\ 
is likely due to the contribution by spectral lines from the CSM in $B$ 
as observed in SNe 2023ixf and 2024ggi,
while the consistent reddening in  \vi\ is 
powered by SCE \citep[e.g.,][]{SW17_SCE}.

\subsection{Progenitor for \kspn}\label{sec:dis_prog}

As shown in \autoref{fig:strip}, the short plateau of \kspn\ may 
be a result of significant (i.e., $\sim$ 3 \msol) envelope stripping of its progenitor,
larger than the $\simlt$ 1 \msol\ mass loss typically expected for a 13 \msol\ ZAMS progenitor 
\citep[e.g.,][]{Beasor_20_massloss, Fuller_CSM_massloss}.
Note that in the case of SN~2018gj, 
a Type II SN with a short plateau which is still longer than that of \kspn,
envelope stripping with a mass loss rate of $\sim$ 0.01 \msol\ yr$^{-1}$ 
has been suggested \citep{SN2018gj}.
Given the large amount of envelope stripping required for \kspn,
we explore a possibility of binary progenitor for enhanced mass loss. 

\citet{Eldridge_18_binary} has conducted SNEC simulations of 
CCSN light curves from binary systems whose progenitor mass, 
initial binary separations ($a$), and mass ratios 
are in the ranges of 5--40 \msol, 
log($a/R_\odot$) =  1--4, and 0.1--0.9, respectively,
using the binary evolution code Binary Population and Spectral Synthesis \citep{BPASS_17, BPASS_18}.
A total of 158 cases from the simulations are for progenitor mass
of 12 \msol\ and explosion energy of 1.0 foe, similar to \kspn.
The two solid green curves in \autoref{fig:strip} are the closest matches 
from the simulations to the observed $V$-band light curve of \kspn,
Although there exist apparent differences between the simulations and observed light curve,
most notably the lack of near-peak excess luminosity 
in the former due to the absence of CSM component in the simulations \citep{Eldridge_18_binary},
the binary progenitors are capable of producing  
a short plateau featured with a fast post-peak decline comparable to that observed in \kspn.
Supernova light curve simulations for binary progenitors with CSM component
are required  to investigate the binary progenitor possibility for \kspn\ 
more thoroughly. 

\section{Summary and Conclusions}\label{sec:conclude}

In this paper, we provide photometric and spectroscopic studies of a luminous 
($M_V$ $\simeq$ --18.41 mag at peak) infant CCSN \kspn\ from redshift $z$ $\simeq$ 0.041,
which was discovered about 15 minutes from the explosion 
estimated by single power law fitting of its early light curves. 
We summarize our results as follows.

\begin{itemize}

    \item  We identify \kspn\ to be a transitional Type II SN between II-P and II-L subtypes. 
    It is a fast-evolving H-rich SN with a short rise time of 4.8 days in the $V$ band 
    together with large post-peak decline rates $s50_V$ $\simeq$ 1.7 \mfif.
    The transitional nature of \kspn\ is confirmed by its rapid evolution and a short plateau, detected only in the $V$ and $i$ band 
    with an OTPd $\sim$ 60 days.
    We estimate the peak bolometric luminosity and \ni56\ mass to be $\sim$ 1.0 $\times$ 10$^{43}$ \ergs\  and 0.042 \msol, respectively. 
    
    \item The presence of CSM interactions in \kspn\ is supported by both 
    photometric and spectroscopic features, including (1) near-peak excess emission
    to SCE, (2) small absorption-to-emission ratio ($a/e$ $\simeq$ 0.076) in \hal,
    (3) tentative presence of high-velocity ($\simgt$ 1.4 $\times$ 10$^4$ \kms) 
    H absorption features, and (4) large ejecta velocities on the order
    of $\sim$ 10$^4$ \kms.
    
    \item  The early ($<$ 10 days since SBO) colour evolution of \kspn\ in 
    \bv\ and \vi\ is suggestive of a delayed SBO near the outer edge of its CSM. 
    We attribute its short-lived blueward evolution in \bv\ during the first six days 
    to contributions by line emission from shocked CSM and its continuous
    reddening in \vi\ to the delayed shock breakout with a small amount of
    unshocked CSM.
    
    \item Our SN light curve simulations with SNEC find 13 \msol\ and 1.05 foe as the best fit
    progenitor mass and explosion energy for \kspn\ as well as 
    CSM mass of 0.73 \msol\ to account for its near-peak excess emission.
    Its short plateau is compatible with a significant
    envelope stripping of $\sim$ 3 \msol. 
    It is worthwhile to investigate the possibility of its progenitor being
    a binary system with light curve simulations optimized for 
    CMS interactions and envelope stripping in binary systems.

    \item  We establish a correlation between post-peak decline rates and CSM mass for Type II SNe,
    showing that CSM mass plays a crucial role in shaping their light curve evolution. 
    This leads to the conclusion that CSM interactions are the primary reason for
    SNe becoming a transitional type between Type II-P and II-L subtypes, including \kspn. 

\end{itemize}

\vskip4pt
\section*{Acknowledgments}
\vskip4pt
We thank Christopher D. Matzner, Xiaofeng Wang, Tanrui Sun, Patrick Sandoval, Conor Ransome and V. Ashley Villar for the many helpful discussions. This research has made use of the KMTNet system operated by the Korea Astronomy and Space Science Institute (KASI) and the data were obtained at three host sites of CTIO in Chile, SAAO in South Africa, and SSO in Australia.
Data transfer from the host site to KASI was supported by the Korea Research Environment Open NETwork (KREONET).
This research was supported by KASI under the R\&D program (Project No. 2025-1-831-02), supervised by the Korea AeroSpace Administration.
This research is also based on observations obtained at the international Gemini-S Observatory, a program of National Science Foundation’s (NSF) NOIRLab, which is managed by the Association of Universities for Research in Astronomy (AURA) under a cooperative agreement with the NSF on behalf of the Gemini Observatory partnership: the NSF (United States), National Research Council (NRC; Canada), Agencia Nacional de Investigaci\'{o}n y Desarrollo (Chile), Ministerio de Ciencia, Tecnolog\'{i}a e Innovaci\'{o}n (Argentina), Minist\'{e}rio da Ci\^{e}ncia, Tecnologia, Inova\c{c}\~{o}es e Comunica\c{c}\~{o}es (MCTI; Brazil), and KASI (Republic of Korea).
The Gemini-S observations were obtained under the Canadian Gemini Office 
(PID:  GS-202?A-Q-???) of the NRC and the K-GMT Science Program (PID: GS-202?B-Q-??)
of KASI and accessed through the Gemini Observatory Archive at NSF’s NOIRLab. D.-S.M. and M.R.D. are supported by Discovery Grants from the Natural Sciences and Engineering Research Council of Canada (NSERC; Nos. RGPIN-2019-06524 and RGPIN-2019-06186, respectively).
D.-S.M. was supported in part by a Leading Edge Fund from the Canadian Foundation for Innovation (CFI; project No. 30951).
M.R.D. was supported in part by the Canada Research Chairs Program and the Dunlap Institute at the University of Toronto.

\facilities{KMTNet,  Gemini-S (GMOS)}

\software{Astropy \citep{astropy:2013, astropy:2018, astropy:2022}, \texttt{HOTPANTS} \citep{HOTPANTS}, \texttt{IRAF} \citep{1993ASPC.52.173T}, pPXF \citep{Cappellari_17_galaxy}, \texttt{scikit-learn} \citep{scikit-learn}, SciPy \citep{2020SciPy-NMeth}, SNAP \citep{Ni_22_method}, Source Extractor \citep{Bertin96}}

\bibliography{ZN7713}{}
\bibliographystyle{aasjournal}
\end{document}